\documentclass[11pt]{article}
\usepackage{amssymb}

\usepackage{graphicx}
\usepackage{amsmath}
\usepackage{makeidx}
\usepackage{indentfirst}

\newcounter{resultnum}[section]

\newcounter{conclusionnum}[section]

\newcounter{conditionnum}[section]

\newcounter{conjecturenum}[section]

\newcounter{examplenum}[section]

\newcounter{exercisenum}[section]

\newcounter{lemmanum}[section]

\newcounter{notationnum}[section]

\newcounter{theoremnum}[section]

\newcounter{definitionnum}[section]

\newcounter{corollarynum}[section]

\newcounter{remarknum}[section]

\newcounter{propositionnum}[section]

\newcounter{acknowledgementnum}[section]

\newcounter{algorithmnum}[section]

\newcounter{axiomnum}[section]

\newcounter{casenum}[section]

\newcounter{claimnum}[section]

\newcounter{summarynum}[section]

\newcounter{problemnum}[section]

\begin{document}

\title{Two--Connection Renormalization and\\
Nonholonomic Gauge Models of Einstein Gravity}
\date{May 5, 2009}
\author{ Sergiu I. Vacaru\thanks{
sergiu.vacaru@uaic.ro, Sergiu.Vacaru@gmail.com;\newline   http://www.scribd.com/people/view/1455460-sergiu }
\\
{\quad} \\
{\small {\textsl{ Science Department, University "Al. I. Cuza" Ia\c si},}
}\\
{\small {\textsl{\ 54 Lascar Catargi  street, 700107, Ia\c si, Romania}} }}
\maketitle

\begin{abstract}
A new framework to perturbative quantum gravity is proposed following the geometry of nonholonomic distributions on (pseudo) Riemannian manifolds. There are considered such distributions and adap\-ted  connections, also completely defined by a metric structure, when gravitational models
with infinite many couplings reduce to two--loop renormalizable effective actions. We use a key result from our partner work arXiv: 0902.0911 that the classical Einstein gravity theory can be reformulated equivalently as a nonholonomic gauge model in the bundle of affine/de Sitter frames on pseudo--Riemannian spacetime. It is proven that (for a class of nonholonomic
constraints and splitting of the Levi--Civita connection into a "renormalizable" distinguished connection, on a base background manifold, and a gauge like distortion tensor, in total space) a nonholonomic differential  renormalization procedure for quantum gravitational fields can be elaborated. Calculation labor is reduced to  one-- and two--loop levels and renormalization group equations for nonholonomic configurations.

\vskip0.2cm



\end{abstract}

\tableofcontents

\section{Introduction}

There were elaborated different perturbative approaches and applications to
quantum gravity of the standard formalism developed in the 1970s with the
aim to quantize arbitrary gauge theories. In the bulk, all those results
where derived using (which proves technically very convenient) the
background field method from the very beginning; see reviews \cite{alvarez}
and, for a short discussion of more recent results, \cite{kiefer}. That
period can be characterized by some final results like that the general
relativity is not renormalizable even at one--loop order, when the coupling
to matter is considered, and neither is pure gravity finite to two loops %
\cite{hooft,goroff,vanven}.

The general conclusion that the Einstein's gravity is perturbatively
nonrenormalizable was long time considered a failure as a quantum field
theory and, as a result, different strategies have been pursued. Here, one
should be mentioned supergravity and string gravity and loop quantum gravity
(see comprehensive summiaries of results and reviews, respectively, in Refs. %
\cite{strmo1,strmo2,strmo3}, which is related in the bulk to the background
field method, and \cite{rov,thiem1,asht}, advocating background independent
and non--perturbative approaches). We also note that some time ago S.
Weinberg suggested that a quantum theory in terms of the metric field may
very well exist, and be renormalizable on a non--perturbative level \cite%
{gomiswein}. That scenario known as ''asymptotic safety'' necessitates an
interacting ultraviolet fixing point for gravity under the renormalization
group (see \cite{bern,niedermaier,lausch,litim,percacci}, for reviews). It
is also similar in spirit to effective field theory approaches to quantum
gravity \cite{donogh,burgess}. \ But unlike a truly fundamental theory, an
effective model cannot be valid up to arbitrary scales. Even substantial
evidence was found for the non--perturbative renormalizability of the
so--called Quantum Einstein Gravity this emerging quantum model is not a
quantization of classical general relativity, see details in \cite%
{reuter1,reuter2}.

However, until today none of the above mentioned approaches has been
accepted to be fully successful: see, for example, important discussions and
critical reviews of results on loop quantum gravity and spin networks \cite%
{nicolai,thiemann}. Not entering into details of those debates, we note that
a number of researches consider that, for instance, the existence of a
semi-classical limit, in which classical Einstein field equations are
supposed to emerge, is still an open problem to be solved in the loop
quantum gravity approach. This is also related to the problem of the
nonrenormalizable ultra--violet divergences that arise in the conventional
perturbative treatment. Finally, there are another questions like it is
possible to succeed, or not, in achieving a 'true' quantum version of full
spacetime covariance, and how to formulate a systematic treatment of
interactions with matter fields etc.

In a series of our recent works \cite{vpla,vfqlf,vqg4,anav}, we proved that
the Einstein gravity theory, redefined in so--called almost K\"{a}hler
variables, can be formally quantized following methods of Fedosov
(deformation) quantization \cite{fed1,fed2,karabeg1}. \ The approach was
derived from the formalism of nonlinear connections and correspondingly
adapted Lagrange--Finsler variables on (pseudo) Riemannian manifolds, as it
was considered in Refs. \cite{ijgmmp,vrflg} (see there details how Finsler
like distributions can be defined on Einstein manifolds which is very
important for constructing generic off--diagonal solutions in general
relativity and elaborating certain new schemes of quantization). We
emphasize that in this article we shall not work with more general classes
of Lagrange--Finsler geometries elaborated in original form on tangent
bundles \cite{ma} but apply the nonholonomic manifold geometric formalism
for our purposes in quantum gravity (with classical and quantum versions of
Einstein manifolds, see also a summary of alternative geometric results on
nonholonomic manifolds in Ref. \cite{bejf}).

Having introduced nonholonomic almost K\"{a}hler variables in Einstein
gravity, the problem of quantization of gravity can be approached \cite%
{vgwgr} following certain constructions for nonholonomic branes and
quantization of a corresponding A--model complexification for gravity as it
was proposed for gauge and topological theories in a recent work by Gukow
and Witten \cite{gukwit}. The nonholonomic canonical symplectic variables
also provide a bridge to nonholonomic versions of Ashtekar--Barbero
variables (and non--perturbative constructions in loop gravity) \cite%
{vloopdq}. They can be applied to more general cases of noncommutative
theories of (gauge and Einstein) gravity \cite{vggr,vncg} and connected to
the theory of nonholonomic/ noncommutative Ricci flows, Perelman functionals
and Dirac operators \cite{vnhrf,vnoncomrf}. The next step in developing the
nonholonomic geometric formalism for classical and quantum gravity theory
consists in a study of general relativity along the lines of
''conventional'' quantum field theory.

Our key idea is to work with an alternative class of metric compatible
linear and nonlinear connections which are completely defined by a metric
tensor and adapted to necessary types of nonholonomic constraints. We shall
use two basic results from the first partner work \cite{vpartn1}:\ 1) The
Einstein gravity theory can be equivalently reformulated in terms of new
variables defined by nonholonomic frames and nonholonomically deformed
connections possessing constant coefficient curvatures\footnote{%
but equivalently to the well known approaches with Levi--Civita, Ashtekar
and various gauge like gravitational connections}.\ 2) The contributions of
distortion tensor (considering deformations from an auxiliary linear
connection to a Levi--Civita one) can be encoded into formal gauge gravity
models\footnote{%
in this work any distinguished connection and relevant distortion tensors
will be completely defined by a metric structure, similarly to the
Levi--Civita connection}.

In this paper, we consider some new perspectives to quantum gravity
following certain methods from the geometry of nonholonomic distributions on
(pseudo) Riemannian manifolds. Our purpose is twofold:

\begin{enumerate}
\item We show that for distinguished connections with constant curvature
coefficients the higher--derivative quadratic terms can be removed by means
of nonholonomic deformations and covariant field redefinitions and vertex
renormalization. This way the theories with infinitely many couplings (like
Einstein gravity, see a detailed discussion in \cite{anselmi}) can be
studied in a perturbative sense also at high energies, despite their
notorious perturbative non--renormalizability.

\item We prove that quantization of nonholonomic distortions of connections
in certain affine/de Sitter frame bundles is possible following standard
perturbative methods for the Yang--Mills theory; for our approach, we shall
use the quantization techniques summarized in Ref. \cite{weinb2}.
\end{enumerate}

Let us now outline the content of this work:

In section 2, we provide some basic formulas, denotations and necessary
results on two--connection variables and nonholonomic gauge models of
Einstein gravity considered in details in Ref. \cite{vpartn1}.

In section 3, we propose a new approach to the problem of renormalization of
gravity theories following geometric constructions with nonholonomic
distributions and alternative connections (to the Levi--Cevita one) also
defined by the same metric structure.

We prove that pure gravity may be two--loop nondivergent, even on shell, but
for an alternative "distinguished" connection, from which various
connections in general relativity theory can be generated by using
corresponding distortion tensors also completely defined by metric tensor.
The method of differential renormalization is generalized on nonholonomic
spaces, for one-- and two--loop calculus, in section 4.

Then, in section 5, we apply this method of quantization to a nonholonomic/
nonlinear gauge gravity theory (classical formulation being equivalent to
Einstein gravity). The one-- and two--loop computations on bundles spaces
enabled with nonholonomic distributions are provided for certain estimations
or running constants and renormalization group equations for nonholonomic
gravitational configurations.

Section 6 is devoted to a summary and conclusions. In Appendix, there are
given some formulas for overlapping divergences in nonholonomic spaces.

\section{Nonholonomic Gauge Models of Einstein Gravity and Two--Connection
Variables}

We consider a four dimensional (pseudo) Riemannian manifold $\mathbf{V}$
with the metric structure parametrized in the form
\begin{eqnarray}
\mathbf{g} &=&g_{\alpha \beta }\mathbf{e}^{\alpha }\otimes \mathbf{e}^{\beta
}=g_{ij}e^{i}\otimes e^{j}+h_{ab}\mathbf{e}^{a}\otimes \mathbf{e}^{b}=
\label{ansatz} \\
\mathbf{\mathring{g}} &=&\mathring{g}_{\alpha \beta }\mathbf{\mathring{e}}%
^{\alpha }\otimes \mathbf{\mathring{e}}^{\beta }=\mathring{g}_{i^{\prime
}j^{\prime }}e^{i^{\prime }}\otimes e^{j^{\prime }}+\mathring{h}_{a^{\prime
}b^{\prime }}\mathbf{\mathring{e}}^{a^{\prime }}\otimes \mathbf{\mathring{e}}%
^{b^{\prime }},  \notag \\
\mathbf{\mathring{e}}^{\alpha ^{\prime }} &=&(e^{i^{\prime }}=dx^{i^{\prime
}},\mathbf{\mathring{e}}^{a^{\prime }}=dy^{a^{\prime }}+\mathring{N}%
_{i^{\prime }}^{a^{\prime }}dx^{i^{\prime }}),  \label{ddifc} \\
\mathbf{e}^{\alpha } &=&(e^{i}=dx^{i},\mathbf{e}^{a}=dy^{a}+N_{i}^{a}dx^{i}),
\label{ddif} \\
\mbox{for \qquad }g_{ij}e_{\ i^{\prime }}^{i}e_{\ j^{\prime }}^{j} &=&%
\mathring{g}_{i^{\prime }j^{\prime }},\ h_{ab}e_{\ a^{\prime }}^{a}e_{\
b^{\prime }}^{b}=\mathring{h}_{a^{\prime }b^{\prime }},\ N_{i}^{a}e_{\
i^{\prime }}^{i}e_{a\ }^{\ a^{\prime }}=\mathring{N}_{i^{\prime
}}^{a^{\prime }},  \label{rediff}
\end{eqnarray}%
with respect to dual bases $\mathbf{e}^{\alpha }$ and $\mathbf{\mathring{e}}%
^{\alpha ^{\prime }},$ for $e^{i}=e_{\ i^{\prime }}^{i}\mathring{e}%
^{i^{\prime }}$ and $\mathbf{e}^{a}=e_{\ a^{\prime }}^{a}\mathbf{\mathring{e}%
}^{a^{\prime }},$ where vierbein coefficients $e_{\ \alpha ^{\prime
}}^{\alpha }=[e_{\ i^{\prime }}^{i},e_{\ a^{\prime }}^{a}]$ are defined by
for any given/prescribed values $g_{\alpha \beta }=[g_{ij},h_{ab}],\mathring{%
g}_{\alpha \beta }=[\mathring{g}_{i^{\prime }j^{\prime }},\mathring{h}%
_{a^{\prime }b^{\prime }}]$ and $\mathring{N}_{i^{\prime }}^{a^{\prime }}.$
For convenience, we can consider constant coefficients $\mathring{g}%
_{i^{\prime }j^{\prime }}$ and $\mathring{h}_{a^{\prime }b^{\prime }}$ and
take $[\mathring{e}^{i^{\prime }},\mathring{e}^{j^{\prime }}]=0.$ The
nonholonomic structure of $\mathbf{V}$ is determined by a nonlinear
connection (N--connection) $\mathbf{N}=N_{i}^{a}(u)dx^{i}\otimes \partial
_{a}.$\footnote{%
Local coordinates on $\mathbf{V}$ are denoted in the form $u^{\alpha
}=(x^{i},y^{a})$ (or, in brief, $u=(x,y)$) where indices of type $%
i,j,...=1,2 $ are formal horizontal/ holonomic ones (h--indices), labeling
h--coordinates, and indices of type $a,b,...=3,4$ are formal
vertical/nonholonomic ones (v--indices), labeling v--coordinates. We may use
'underlined' indices ($\underline{\alpha }=(\underline{i},\underline{a}),%
\underline{\beta }=(\underline{i},\underline{b}),...$), for local coordinate
bases $e_{\underline{\alpha }}=\partial _{\underline{\alpha }}=(\partial _{%
\underline{i}},\partial _{\underline{a}}),$ equivalently $\partial /\partial
u^{\underline{\alpha }}=(\partial /\partial x^{\underline{i}},\partial
/\partial y^{\underline{a}});$ for dual coordinate bases we shall write \ $%
e^{\underline{\alpha }}=du^{\underline{\alpha }}=(e^{\underline{i}}=dx^{%
\underline{i}},e^{\underline{a}}=dx^{\underline{a}}).$ There are also
considered primed indices ($\alpha ^{\prime }=(i^{\prime },a^{\prime
}),\beta ^{\prime }=(j^{\prime },b^{\prime }),...$), with double primes etc,
for other local abstract/coordinate bases, for instance, $e_{\alpha ^{\prime
}}=(e_{i^{\prime }},e_{a^{\prime }}),e^{\alpha ^{\prime }}=(e^{i^{\prime
}},e^{a^{\prime }})$ and $e_{\alpha ^{\prime \prime }}=(e_{i^{\prime \prime
}},e_{a^{\prime \prime }}),e^{\alpha ^{\prime \prime }}=(e^{i^{\prime \prime
}},e^{a^{\prime \prime }}),$ where $i^{\prime },i^{\prime \prime },=1,2...$
and $a^{\prime },a^{\prime \prime }=3,4.$} Such a manifold is called
N--anholonomic. On adopted system of notations and details on nonholonomic
manifolds and N--connection geometry and applications in modern gravity, we
cite our first partner work \cite{vpartn1} and Refs. \cite{ijgmmp,vrflg}.%
\footnote{%
On a manifold $\mathbf{V,}$ we can fix any type of coordinate and frame
(equivalently, vierbein/ tetradic) and nonholonomic, in our case,
N--connection, structures; this will result in different types of
coefficients $\mathring{N}_{i^{\prime }}^{a^{\prime }},N_{i^{\prime
}}^{a^{\prime }}$ and $e_{\ i^{\prime }}^{i},e_{a\ }^{\ a^{\prime }},$
respectively, in formulas (\ref{ddifc}),(\ref{ddif}) and (\ref{rediff}), for
any given metric $\mathbf{g}=\{g_{\alpha \beta }\}$ and fixed values (for
instance, constant) $\mathring{g}_{i^{\prime }j^{\prime }}$ and $\mathring{h}%
_{a^{\prime }b^{\prime }}.$}

On a N--anholonomic $\mathbf{V,}$ we can construct an infinite number of
(linear) distinguished connections, d--connections, $\mathbf{D}=\{\mathbf{%
\Gamma }_{\ \alpha \beta }^{\gamma }\}$ which are adapted to a chosen
N--connection structure $\mathbf{N}$ (i.e. the N--connection h-- and
v--splitting is preserved under parallelism) and metric compatible, $\mathbf{%
Dg}=0.$ There is a subclass of such d--connections $\ ^{\mathbf{g}}\mathbf{D}
$ $\ $when their coefficients $\ ^{\mathbf{g}}\mathbf{\Gamma }_{\ \alpha
\beta }^{\gamma }$ are uniquely determined by the coefficients of $\ \mathbf{%
g=\mathring{g}}$ following a geometric principle. We can work equivalently
with the Levi--Civita connection $\ ^{\mathbf{g}}\nabla =\{\ _{\shortmid }^{%
\mathbf{g}}\Gamma _{\ \beta \gamma }^{\alpha }\}$ and any $\ ^{\mathbf{g}}%
\mathbf{\Gamma }_{\ \alpha \beta }^{\gamma }$ related by a distortion
relation $\ _{\shortmid }^{\mathbf{g}}\Gamma _{\ \beta \gamma }^{\alpha }=\
^{\mathbf{g}}\mathbf{\Gamma }_{\ \alpha \beta }^{\gamma }+\ _{\shortmid }^{%
\mathbf{g}}\mathbf{Z}_{\ \beta \gamma }^{\alpha }$ because the
distortion/torsion tensor $\ _{\shortmid }^{\mathbf{g}}\mathbf{Z}_{\ \beta
\gamma }^{\alpha }$ is also completely defined by the coefficients $%
g_{\alpha \beta }$ for any prescribed values $N_{i}^{a}.$\footnote{%
We used the left label ''$\mathbf{g"}$ in order to emphasize that certain
values are defined by the metric structure. Such constructions do not depend
explicitly on the type of nonholonomic distribution, for instance, we can
consider any type of $2+2$ distributions (i.e. we work with well defined
geometric objects, not depending on a particular choice of coordinate/ frame
systems, even the constructions are adapted to a fixed nonholonomic
structure). A general d--connection $\mathbf{\Gamma }_{\ \alpha \beta
}^{\gamma }$ is not defined by a metric tensor. For simplicity, in this
work, we shall work only with metric compatible d--connections.}

For our nonholonomic constructions in classical and quantum gravity, a
crucial role is played by the \textit{Miron's procedure }(on applications in
modern gravity and generalizations, see discussions in Refs.\cite%
{vpartn1,vnslms,vgwgr} and the original results, for Lagrange--Finsler
spaces, \cite{ma}). This procedure allows us to compute the set of
d--connections $\{\mathbf{D}\}$ satisfying the conditions $\mathbf{D}_{%
\mathbf{X}}\mathbf{g=0}$ for a given $\mathbf{g.}$ The components of any
such $\mathbf{D=}\left( L_{\ jk}^{i},L_{\ bk}^{a},C_{\ jc}^{i},\ C_{\
bc}^{a}\right) $ are given by formulas%
\begin{eqnarray}
L_{\ jk}^{i} &=&\widehat{L}_{jk}^{i}+\ ^{-}O_{km}^{ei}\mathbf{Y}%
_{ej}^{m\,},\ L_{\ bk}^{a}=\widehat{L}_{bk}^{a}+\ ^{-}O_{bd}^{ca}\mathbf{Y}%
_{ck}^{d\,},  \label{mcdc} \\
C_{\ jc}^{i} &=&\widehat{C}_{jc}^{i}+\ ^{+}O_{jk}^{mi}\mathbf{Y}%
_{mc}^{k\,},\ C_{\ bc}^{a}=\widehat{C}_{bc}^{a}+\ ^{+}O_{bd}^{ea}\mathbf{Y}%
_{ec}^{d\,},  \notag
\end{eqnarray}%
where
\begin{equation*}
\ ^{\pm }O_{jk}^{ih}=\frac{1}{2}(\delta _{j}^{i}\delta _{k}^{h}\pm
g_{jk}g^{ih}),\ ^{\pm }O_{bd}^{ca}=\frac{1}{2}(\delta _{b}^{c}\delta
_{d}^{a}\pm g_{bd}g^{ca})
\end{equation*}%
are the so--called the Obata operators and $\widehat{\mathbf{\Gamma }}_{\
\alpha \beta }^{\gamma }=\left( \widehat{L}_{jk}^{i},\widehat{L}_{bk}^{a},%
\widehat{C}_{jc}^{i},\widehat{C}_{bc}^{a}\right) ,$ with
\begin{eqnarray}
\widehat{L}_{jk}^{i} &=&\frac{1}{2}g^{ir}\left( \mathbf{e}_{k}g_{jr}+\mathbf{%
e}_{j}g_{kr}-\mathbf{e}_{r}g_{jk}\right) ,  \label{candcon} \\
\widehat{L}_{bk}^{a} &=&e_{b}(N_{k}^{a})+\frac{1}{2}g^{ac}\left( \mathbf{e}%
_{k}g_{bc}-g_{dc}\ e_{b}N_{k}^{d}-g_{db}\ e_{c}N_{k}^{d}\right) ,  \notag \\
\widehat{C}_{jc}^{i} &=&\frac{1}{2}g^{ik}e_{c}g_{jk},\ \widehat{C}_{bc}^{a}=%
\frac{1}{2}g^{ad}\left( e_{c}g_{bd}+e_{c}g_{cd}-e_{d}g_{bc}\right) ,  \notag
\end{eqnarray}%
is the canonical d--connection uniquely defined by the coefficients of
d--metric $\mathbf{g=}[g_{ij},g_{ab}]$ and N--connection $\mathbf{N}%
=\{N_{i}^{a}\}$ in order to satisfy the conditions $\widehat{\mathbf{D}}_{%
\mathbf{X}}\mathbf{g=0}$ and $\widehat{T}_{\ jk}^{i}=0$ and $\widehat{T}_{\
bc}^{a}=0$ but with general nonzero values for $\widehat{T}_{\ ja}^{i},%
\widehat{T}_{\ ji}^{a}$ and $\widehat{T}_{\ bi}^{a},$ see component formulas
in \cite{vpartn1,ma}. In formulas (\ref{mcdc}), the d--tensors $\mathbf{Y}%
_{ej}^{m\,},\mathbf{Y}_{mc}^{k\,},\mathbf{Y}_{ck}^{d\,}$ and $\mathbf{Y}%
_{ec}^{d\,}$ parametrize the set of metric compatible d--connections, with a
metric $\mathbf{g,}$ on a N--anholonomic manifold $\mathbf{V}.$ Prescribing
any values of such d--tensors (following certain geometric/ physical
arguments; in particular, we can take some zero, or non--zero, constants),
we get a metric compatible d--connection $\ ^{\mathbf{g}}\mathbf{D}$ (\ref%
{mcdc}) completely defined by a (pseudo) Riemannian metric $\mathbf{g}$ (\ref%
{ansatz}). Such d--connections can be chosen in different forms for
different quantization/ renormalization procedures in modern quantum gravity%
\footnote{%
see, for instance, applications of the geometry of nonholonomic
distributions and nonlinear connections in Refs. \cite%
{vpla,vfqlf,vqg4,anav,vgwgr,vloopdq}} (on the two--connection perturbative
method, see next sections in this work). For simplicity, we shall consider
that we fix a nonholonomic configuration of frames and linear connections on
a spacetime manifold $V,$ and lifts of fundamental geometric objects
(metrics, connections, tensors, physical fields etc) on total spaces of some
bundles on $V,$ if prescribe certain constant (for simplicity), or tensor
fields for ${}$--fields in formulas (\ref{mcdc}).

It is possible to construct nonholonomic lifts of any connections $\
_{\shortmid }^{\mathbf{g}}\Gamma _{\ \beta \gamma }^{\alpha }$ and \ $\ ^{%
\mathbf{g}}\mathbf{\Gamma }_{\ \alpha \beta }^{\gamma },$ and related
distortion tensors $\ _{\shortmid }^{\mathbf{g}}\mathbf{Z}_{\ \beta \gamma
}^{\alpha },$ into the bundle of affine/ de Sitter frames on a
N--anholonomic spacetime $\mathbf{V}$ (see details in Section 3 of Ref. \cite%
{vpartn1}). In our approach, the de Sitter nonlinear gauge gravitational
theory is constructed from the coefficients of a d--metric $\mathbf{g}$ and
N--connection $\mathbf{N}$ in a form when the Einstein equations on the base
nonholonomic spacetime are equivalent to the Yang--Mills equations in a
total space enabled with induced nonholonomic structure. We choose in the
total de Sitter nonholonomic bundle a d--connection $\ ^{\mathbf{g}}\mathbf{%
\Gamma }$ which with respect respect to nonholonomic frames of type (\ref%
{ddif}), and their duals, is determined by a d--connection $\ ^{\mathbf{g}}%
\mathbf{\Gamma }_{\ \beta \gamma }^{\alpha },$
\begin{equation}
\ ^{\mathbf{g}}\mathbf{\Gamma }=\left(
\begin{array}{cc}
\ ^{\mathbf{g}}\mathbf{\Gamma }_{\quad \beta ^{\prime }}^{\alpha ^{\prime }}
& l_{0}^{-1}\mathbf{e}^{\alpha ^{\prime }} \\
l_{1}^{-1}\mathbf{e}_{\beta ^{\prime }} & 0%
\end{array}%
\right) ,  \label{conds}
\end{equation}%
where
\begin{equation}
\ ^{\mathbf{g}}\mathbf{\Gamma }_{\quad \beta ^{\prime }}^{\alpha ^{\prime
}}=\ ^{\mathbf{g}}\mathbf{\Gamma }_{\quad \beta ^{\prime }\mu }^{\alpha
^{\prime }}\mathbf{e}^{\mu },  \label{condsc}
\end{equation}%
for
\begin{equation}
\ ^{\mathbf{g}}\Gamma _{\quad \beta ^{\prime }\mu }^{\alpha ^{\prime }}=%
\mathbf{e}_{\alpha }^{~\alpha ^{\prime }}\mathbf{e}_{\quad \beta ^{\prime
}}^{\beta }\ ^{\mathbf{g}}\mathbf{\Gamma }_{\quad \beta \mu }^{\alpha }+%
\mathbf{e}_{\alpha }^{~\alpha ^{\prime }}\mathbf{e}_{\mu }(\mathbf{e}_{\quad
\beta ^{\prime }}^{\alpha }),\   \label{coeff3}
\end{equation}%
with $\mathbf{e}^{\alpha ^{\prime }}=\mathbf{e}_{\mu }^{~\alpha ^{\prime }}%
\mathbf{e}^{\mu }$ and $l_{0}$ and $l_{1}$ being dimensional constants. The
indices $\alpha ^{\prime },\beta ^{\prime }$ take values in the typical
fiber/ de Sitter space.\footnote{%
In a similar form we can elaborate certain geometric constructions for
nonholonoic affine frame bundles if we chose $^{\mathbf{g}}\mathbf{\Gamma }%
=\left(
\begin{array}{cc}
\ ^{\mathbf{g}}\mathbf{\Gamma }_{\quad \beta ^{\prime }}^{\alpha ^{\prime }}
& l_{0}^{-1}\mathbf{e}^{\alpha ^{\prime }} \\
0 & 0%
\end{array}%
\right) $ but this results in formal 'non-variational' gauge models because
of degenerated Killing forms. Geometrically, this is not a problem and, in
both cases (for instance, for the affine and de Sitter frame bundles) we can
work with well--defined nonholonomic structures and geometric objects in
total bundle, considering necessary auxiliar tensor fields and constants
defining a class of N--connections, when the projections of nonholonomic
Yang--Mills equations on a spacetime base will be equivalent to the Einstein
equations.} We emphasize that because of nonholonomic structure on the base
and total spaces, the coefficients of $\ ^{\mathbf{g}}\mathbf{\Gamma }$ are
subjected to nonlinear nonholonomic transformations laws under group/ frame/
coordinate transforms and nonhlonomic deformations, see explicit formulas in %
\cite{vpartn1}. If we take the limit $l_{1}^{-1}\rightarrow 0$, we get a
d--connection for the affine frame bundle (with degenerated fiber metric)
which allows to project the connection 1--form just in a d--connection on
the base, when the constructions can be performed to be equivalent to the
Einstein gravity. For $l_{1}=l_{0},$ we shall develop a de Sitter model with
nondegenerate fiber metric. For simplicity, we shall work only with a
nonholonomic de Sitter frame model of nonholonomic gauge gravity.

The matrix components of the curvature of the d--connection (\ref{conds}),
\begin{equation*}
\ ^{\mathbf{g}}\mathcal{R}=d\ ^{\mathbf{g}}\mathbf{\Gamma }-\ ^{\mathbf{g}}%
\mathbf{\Gamma }\wedge \ ^{\mathbf{g}}\mathbf{\Gamma },
\end{equation*}%
can be parametrized in an invariant 4+1 form
\begin{equation}
\ \ ^{\mathbf{g}}\mathcal{R}=\left(
\begin{array}{cc}
\mathcal{R}_{\quad \beta ^{\prime }}^{\alpha ^{\prime }}+l_{0}^{-1}\pi
_{\beta ^{\prime }}^{\alpha ^{\prime }} & l_{0}^{-1}\mathcal{T}^{\alpha
^{\prime }} \\
l_{0}^{-1}\mathcal{T}^{\beta ^{\prime }} & 0%
\end{array}%
\right) ,  \label{curvs}
\end{equation}%
where%
\begin{eqnarray*}
\pi _{\beta ^{\prime }}^{\alpha ^{\prime }} &=&\mathbf{e}^{\alpha ^{\prime
}}\wedge \mathbf{e}_{\beta ^{\prime }},~\mathcal{T}^{\beta ^{\prime }}=\frac{%
1}{2}\ ^{\mathbf{g}}\mathbf{T}_{\quad \mu \nu }^{\beta ^{\prime }}\delta
u^{\mu }\wedge \delta u^{\nu } \\
\mathcal{R}_{\quad \beta ^{\prime }}^{\alpha ^{\prime }} &=&\frac{1}{2}%
\mathcal{R}_{\quad \beta ^{\prime }\mu \nu }^{\alpha ^{\prime }}\delta
u^{\mu }\wedge \delta u^{\nu },\ \mathcal{R}_{\quad \beta ^{\prime }\mu \nu
}^{\alpha ^{\prime }}=\mathbf{e}_{~\beta ^{\prime }}^{\beta }\mathbf{e}%
_{\alpha }^{\ \alpha ^{\prime }}\ ^{\mathbf{g}}\mathbf{R}_{\quad \beta _{\mu
\nu }}^{\alpha },\
\end{eqnarray*}%
when the torsion, $\ ^{\mathbf{g}}\mathbf{T}_{\quad \mu \nu }^{\beta
^{\prime }},$ and curvature, $\ ^{\mathbf{g}}\mathbf{R}_{\quad \beta {\mu
\nu }}^{\alpha },$ tensors are computed for the connection 1--form (\ref%
{condsc}). The constant $l_{0}$ in (\ref{conds}) and (\ref{curvs}) and
constants $l^{2}=2l_{0}^{2}\lambda ,\lambda _{1}=-3/l_{0}$ considered in
Ref. \cite{vpartn1} do not characterize certain additional gravitational
high curvature and/or torsion interactions like in former gauge like gravity
theories \cite{pd1,pd2,tseytl,hehl,sard,vgon,vncg1,vd,vncg}, but define the
type of nonholonomic constraints on de Sitter/affine bundles which are used
for an equivalent lift in a total bundle space of the Einstein equations on a base spacetime
manifold. Prescribing certain values of such constant is equivalent to a
particular choice of $\mathbf{Y}$--fields in formulas (\ref{mcdc}) in order
to fix a nonholonomic configuration\footnote{%
see discussion of Miron's procedure in Section 2 of Ref. \cite{vpartn1}}.
Different values of such nonholonomy constants and additional tensor fields
parametrize various type of N--adapted metric compatible linear connections
and effective gauge gravitational models into which the geometric and
physical information of classical Einstein gravity can be encoded. For our
purposes in developing a method of perturbative quantization of gravity, it
is enough to fix a convenient set of constant $\mathbf{Y}$--fields, which
under quantization will run following certain renormalization group
equations, see below the end of Section 5.

   We can fix such a nonholonomic distribution (and
nonholonomic frames) when $\mathbf{g=\mathring{g}}$ induces a canonical
d--connection $\widehat{\mathbf{\mathring{\Gamma}}}_{\ \alpha ^{\prime
}\beta ^{\prime }}^{\gamma ^{\prime }}=(0,\widehat{\mathring{L}}_{\
b^{\prime }k^{\prime }}^{a^{\prime }},0,0)$ $= const,$ with constant curvature
coefficients
\begin{equation}
\ \widehat{\mathbf{\mathring{R}}}_{\ \beta ^{\prime }\gamma ^{\prime }\delta
^{\prime }}^{\alpha ^{\prime }}=(0,\widehat{\mathring{R}}_{~b^{\prime
}j^{\prime }k^{\prime }}^{a^{\prime }}=\ \widehat{\mathring{L}}_{\ b^{\prime
}j^{\prime }}^{c^{\prime }}\ \widehat{\mathring{L}}_{\ c^{\prime }k^{\prime
}}^{a^{\prime }}-\ \widehat{\mathring{L}}_{\ b^{\prime }k^{\prime
}}^{c^{\prime }}\ \widehat{\mathring{L}}_{\ c^{\prime }j^{\prime
}}^{a^{\prime }},0,0,0,0),  \label{auxdcurv}
\end{equation}%
with respect to a class of N--adapted frames.\footnote{%
d--connections with constant curvature matrix coefficients were introduced
with the aim to encode classical Einstein equations into nonholonomc
solitonic hierarchies \cite{vacap}, see also \cite{ancv} and, on the
procedure of metrization and parametrization of metric compatible
d--connections on general holonomic manifolds/bundle spaces enabled with
symmetric, or nonsymmetric, metrics, \cite{vnslms}} The corresponding
distortion of the Levi--Civita connection with respect to $\widehat{\mathbf{%
\mathring{\Gamma}}}_{\ \alpha ^{\prime }\beta ^{\prime }}^{\gamma ^{\prime
}} $ is written in the form $\ _{\shortmid }\Gamma _{\ \beta \gamma
}^{\alpha }=\widehat{\mathbf{\mathring{\Gamma}}}_{\ \beta \gamma }^{\alpha
}+\ _{\shortmid }\widehat{\mathbf{\mathring{Z}}}_{\ \beta \gamma }^{\alpha
}. $ The related distortions in the total space of nonholonomic fiber
bundles are $\ _{\shortmid }^{\mathbf{g}}\overline{\Gamma }=\ \overline{%
\mathbf{\mathring{\Gamma}}}+\ _{\shortmid }\ \overline{\mathbf{\mathring{Z}}}
$ \ and $\ \ _{\shortmid }^{\mathbf{g}}\overline{\mathcal{R}}=\ \overline{%
\mathcal{\mathring{R}}}+\ _{\shortmid }\overline{\mathcal{\mathring{Z}}}.$

For a four dimensional (pseudo) Riemannian base $\mathbf{V,}$ one could be
maximum eight nontrivial components $\widehat{\mathring{L}}_{\ b^{\prime
}k^{\prime }}^{a^{\prime }}.$ We can prescribe such a nonholonomic
distribution with some nontrivial values $\widehat{\mathring{L}}_{\
b^{\prime }j^{\prime }}^{c^{\prime }}$ when%
\begin{equation}
\widehat{\mathring{R}}_{~b^{\prime }j^{\prime }k^{\prime }}^{a^{\prime }}=\
\widehat{\mathring{L}}_{\ b^{\prime }j^{\prime }}^{c^{\prime }}\ \widehat{%
\mathring{L}}_{\ c^{\prime }k^{\prime }}^{a^{\prime }}-\ \widehat{\mathring{L%
}}_{\ b^{\prime }k^{\prime }}^{c^{\prime }}\ \widehat{\mathring{L}}_{\
c^{\prime }j^{\prime }}^{a^{\prime }}=0.  \label{looptwod}
\end{equation}%
Here it should be emphasized that in order to perform a so--called
two--connection geometric renormalization of two--loop Einstein gravity, it
is possible to consider any metric compatible d--connection $\ ^{\mathbf{g}}%
\mathbf{\Gamma }_{\quad \beta \mu }^{\alpha }$ with corresponding curvature
d--tensor $\ ^{\mathbf{g}}\mathbf{R}_{\quad \beta _{\mu \nu }}^{\alpha }$
satisfying the condition
\begin{equation}
\ ^{\mathbf{g}}\mathbf{R}_{\alpha \beta }^{\quad \mu \nu }\ ^{\mathbf{g}}%
\mathbf{R}_{_{\mu \nu }}^{\quad \gamma \tau }\ ^{\mathbf{g}}\mathbf{R}%
_{_{\gamma \tau }}^{\quad \gamma \tau }=0,  \label{constr}
\end{equation}%
see below Section \ref{stlc}. Such a condition is satisfied by any $\widehat{%
\mathbf{\mathring{\Gamma}}}_{\ \alpha ^{\prime }\beta ^{\prime }}^{\gamma
^{\prime }}$ with prescribed constant coefficients of type $\widehat{%
\mathring{L}}_{\ b^{\prime }k^{\prime }}^{a^{\prime }}$ with vanishing $%
\widehat{\mathring{R}}_{~b^{\prime }j^{\prime }k^{\prime }}^{a^{\prime }}$ (%
\ref{looptwod}), or with such prescribed constants when certain contractions
of this d--tensor are constant/zero.\footnote{%
The geometric properties of curvature and Weyl d--tensors for a
d--connection are very different from those of usual tensors and linear
connections. Even the coefficients of a d--tensor may vanish with respect to
a particular class of nonholonomic distributions, the real spacetime may be
a general (pseudo) Riemannian one with nontrivial curvature of the
Levi--Civita connection and nonzero associated/induced nonholonomically
d--torsions, nonholonomy coefficients and curvature of N--connection.}

Choosing $\ \overline{\mathcal{\mathring{R}}}=0,$ we can write the gauge
like gravitational equations, the equivalent of the Einstein equations on $%
\mathbf{V},$ in a simplified form,%
\begin{equation}
d\left( \ast \ _{\shortmid }\overline{\mathcal{\mathring{Z}}}\right) +\
_{\shortmid }\ \overline{\mathbf{\mathring{Z}}}\wedge \left( \ast \ \
_{\shortmid }\overline{\mathcal{\mathring{Z}}}\right) -\left( \ast \ \
_{\shortmid }\overline{\mathcal{\mathring{Z}}}\right) \wedge \ \ _{\shortmid
}\ \overline{\mathbf{\mathring{Z}}}=-\ _{\shortmid }\overline{\mathcal{%
\mathring{J}}},  \label{efymeq1}
\end{equation}%
where the nonholonomically deformed source is%
\begin{equation*}
\ _{\shortmid }\overline{\mathcal{\mathring{J}}}=\ _{\shortmid }^{\mathbf{g}}%
\overline{\mathcal{J}}+\ \overline{\mathbf{\mathring{\Gamma}}}\wedge \left(
\ast \ _{\shortmid }\overline{\mathcal{\mathring{Z}}}\right) -\left( \ast \
_{\shortmid }\overline{\mathcal{\mathring{Z}}}\right) \wedge \ \overline{%
\mathbf{\mathring{\Gamma}}},
\end{equation*}%
$\ _{\shortmid }^{\mathbf{g}}\overline{\mathcal{J}}$ determined by the
energy--momentum tensor in general relativity and $\ _{\shortmid }\overline{%
\mathcal{\mathring{Z}}}$ contains the same geometric/physical information as
the curvature and Ricci tensor of the Levi--Civita connection $\ _{\shortmid
}\Gamma _{\ \beta \gamma }^{\alpha }.$ Such formulas were derived in Section
4.3 of Ref. \cite{vpartn1}, see there explicit component formulas for $\
\overline{\mathbf{\mathring{\Gamma}}},\ \ _{\shortmid }\ \overline{\mathbf{%
\mathring{Z}}},\ _{\shortmid }\overline{\mathcal{\mathring{Z}}}$ and $\
_{\shortmid }\overline{\mathcal{\mathring{J}}}.$

Finally, we note that from formulas (\ref{auxdcurv}) and (\ref{looptwod})
one follows that
\begin{equation}
\ \widehat{\mathbf{\mathring{R}}}_{\ \beta ^{\prime }\gamma ^{\prime }}=\
\widehat{\mathbf{\mathring{R}}}_{\ \beta ^{\prime }\gamma ^{\prime }\alpha
^{\prime }}^{\alpha ^{\prime }}=0.  \label{vacdeinst}
\end{equation}%
This is similar to the vacuum Einstein equations, but because the Ricci
d--tensor $\ \widehat{\mathbf{\mathring{R}}}_{\ \beta ^{\prime }\gamma
^{\prime }}$ is constructed for a ''nonholonomic'' d--connection$\ \overline{%
\mathbf{\mathring{\Gamma}}}$ this structure is not trivial even its
curvature d--tensor may vanish for certain parametrizations. In such cases,
 a part of''gravitational'' degrees of freedom are encoded into the nonholonomy
coefficients and associated N--connection structure. How to construct
nontrivial exact solutions of (\ref{vacdeinst}) was considered in a series
of our works, see reviews \cite{ijgmmp,vrflg,vncg}. In a more general case,
we can work with nonholonomic configurations when $\ \widehat{\mathbf{%
\mathring{R}}}_{\ \beta ^{\prime }\gamma ^{\prime }}=const,$ for some
nonzero values.

\section{Models of Nonholonomic Gravity and Renormalization}

\label{stlc}It has been shown by explicit computations that in terms of the
Levi--Civita connection the gravity with the Einstein--Hilber action gives
rise to a finite one--loop model only in the absence of both matter fields
and a cosmological constant \cite{hooft}. Similar computations result in a
more negative result that perturbative quantum gravity, with the same
connection, deverges in two--loop order \cite{goroff}, see further results
and review in \cite{vanven,alvarez}. The goal of this section is to prove
that working with nonholonomic distrubutions and alternative d--connections,
the problem of renormalization of gravity theories can be approached in a
different form when certain formal renormalization schemes can be elaborated.

\subsection{Two--loop quantum divergences for nonholonomic models of
gravi\-ty}

We shall analyze the one-- and two--loop divergences of a gravitational
model when the Levi--Civita connection $\ ^{\mathbf{g}}\nabla =\{\
_{\shortmid }^{\mathbf{g}}\Gamma _{\ \beta \gamma }^{\alpha }\}$ is
substituted by an alternative metric compatible d--connection $\ ^{\mathbf{g}%
}\mathbf{D}=$ $\{\ ^{\mathbf{g}}\mathbf{\Gamma }_{\ \beta \gamma }^{\alpha
}\} $ also completely defined by the same metric structure $\mathbf{g}$ in
such a form that the divergences can be eliminated by imposing nonholonomic
constraints.

\subsubsection{One--loop computations}

Let us consider a Lagrange density
\begin{equation}
\ ^{0}\mathcal{L}=-\frac{1}{2\kappa ^{2}}\sqrt{\mathbf{g}}\left( \ ^{\mathbf{%
g}}\overleftrightarrow{\mathbf{R}}-2\Lambda \right)  \label{ld0}
\end{equation}%
where $\kappa ^{2}$ and $\Lambda $ are defined respectively by gravitational
and cosmological constants and the scalar curvature of $\ ^{\mathbf{g}}%
\mathbf{D}$ is
\begin{equation*}
\ ^{\mathbf{g}}\overleftrightarrow{\mathbf{R}}\doteqdot \mathbf{g}^{\alpha
\beta }\ ^{\mathbf{g}}\mathbf{R}_{\alpha \beta }=g^{ij}\ ^{\mathbf{g}%
}R_{ij}+h^{ab}\ ^{\mathbf{g}}S_{ab}=\ ^{\mathbf{g}}\overrightarrow{R}+\ ^{%
\mathbf{g}}\overleftarrow{S},
\end{equation*}
see formula (A.5) in Appendix to \cite{vpartn1}. For simplicity, we restrict
our one-- and two--loop analysis only to nonholonomic vacuum configurations
with $\Lambda =0.$

A one--loop computation similar to that of 't Hooft and Veltman \cite{hooft}%
, see also details in review \cite{alvarez}, but for a background field
method with metric $\mathbf{g}$ and d--connection $\ ^{\mathbf{g}}\mathbf{D,}
$ results in this divergent part of the one--loop effective action for pure
nonholonomic gravity model,%
\begin{equation*}
\mathbf{\Gamma }_{\infty }^{(1)}=\int \delta ^{4}u\sqrt{\mathbf{g}}\left(
a_{1}\ ^{\mathbf{g}}\overleftrightarrow{\mathbf{R}}^{2}+a_{2}\ ^{\mathbf{g}}%
\mathbf{R}_{\alpha \beta }\ ^{\mathbf{g}}\mathbf{R}^{\alpha \beta }+a_{3}\ ^{%
\mathbf{g}}\mathbf{R}_{\alpha \beta \gamma \tau }\ ^{\mathbf{g}}\mathbf{R}%
^{\alpha \beta \gamma \tau }\right) ,
\end{equation*}%
where $a_{1},a_{2}$ and $a_{3}$ are constant. For a given metric structure $%
\mathbf{g,}$ we can always define  a N--connection splitting and
construct  a d--connection $\ ^{\mathbf{g}}\mathbf{\Gamma }_{\ \beta
\gamma }^{\alpha }=\widehat{\mathbf{\mathring{\Gamma}}}_{\ \beta \gamma
}^{\alpha }$ with constant curvature coefficients $\ $when $\ \widehat{%
\mathbf{\mathring{R}}}_{\ \beta ^{\prime }\gamma ^{\prime }\delta ^{\prime
}}^{\alpha ^{\prime }}=0$ (\ref{auxdcurv}) $\ \widehat{\mathbf{\mathring{R}}}%
_{\ \beta ^{\prime }\gamma ^{\prime }}=0$ (\ref{vacdeinst}). In such a case $%
\mathbf{\Gamma }_{\infty }^{(1)}[\widehat{\mathbf{\mathring{\Gamma}}}_{\
\beta \gamma }^{\alpha }]=0$ which holds true for a corresponding class of
nonhlonomic transform and d--connections even the metric $\mathbf{g}$ is a
solution of certain non--vacuum Einstein equations for the Levi--Civita
connection $\nabla $ and nontrivial source of matter fields.

For distortions $\ _{\shortmid }\Gamma _{\ \beta \gamma }^{\alpha }=\widehat{%
\mathbf{\mathring{\Gamma}}}_{\ \beta \gamma }^{\alpha }+\ _{\shortmid }%
\widehat{\mathbf{\mathring{Z}}}_{\ \beta \gamma }^{\alpha },$ we construct a
one--loop finite $\ _{\shortmid }\Gamma _{\ \beta \gamma }^{\alpha }$ if $%
\widehat{\mathbf{\mathring{\Gamma}}}_{\ \beta \gamma }^{\alpha }$ is made
finite by certain nonholonomic transforms and $\ _{\shortmid }\widehat{%
\mathbf{\mathring{Z}}}_{\ \beta \gamma }^{\alpha }$ is renormalized
following some standard methods in gauge theory (they will include also
possible contributions of matter fields). So, we can eliminate the one--loop
divergent part for a corresponding class of metric compatible
d--connections, by corresponding nonholonomic frame deformations.

\subsubsection{Two--loop computations}

The geometry of d--connections adapted to a N--connection structure is more
rich than that of linear connections on manifolds. There are different
conservation laws for d--connections and the derived Ricci and Riemannian
d--tensors contain various types of h-- and v--components inducing different
types of invariants etc (for instance, even in the holonomic case, the
extension of 't Hooft's theorems to two--loop order, for renormalizable
interactions, request an analysis of some 50 invariants, see review \cite%
{alvarez}), see details in Refs. \cite{ijgmmp,vrflg} and, for bundle spaces
and Lagrange--Finsler geometry, \cite{ma}.

Nevertheless, for nonholonomic geometries induced on (pseudo) Riemannian
manifolds and lifted equivalently on bundle spaces, the background field
method works in a similar case both for the Levi--Civita and any metric
compatible d--connection all induces by the same metric structure. From
formal point of view, we have to take $\ _{\shortmid }\Gamma _{\ \beta
\gamma }^{\alpha }\rightarrow \ $ $^{\mathbf{g}}\mathbf{\Gamma }_{\ \beta
\gamma }^{\alpha }$ and follow the same formalism but taking into account
such properties that, for instance, the Ricci d--tensor \ is nonsymmetric,
in general, and that there are additional h-- and v--components with
different transformation laws and invariant properties. We provided such
details for locally anisotropic gravity models, Lagrange--Finsler like and
more general ones, obtained in certain limits of (super) string theory \cite%
{vstr1,vstr2}.\footnote{%
Those results can be redefined equivalently for certain limits to the
Einstein gravity theory and string generalizations considering that the h--
and v--components are not for tangent or vector bundles, but some respective
holonomic and nonholonomic variables on Einstein manifolds.}

In abstract form, the result of a N--adapted background field calculus for
the only gauge--independent coefficient is
\begin{equation*}
\mathbf{\Gamma }_{\infty }^{(2)}=\frac{209}{2880(4\pi )^{4}}\frac{1}{%
\epsilon }\int \delta ^{4}u\sqrt{\mathbf{g}}\ \ ^{\mathbf{g}}\mathbf{R}%
_{\alpha \beta }^{\quad \mu \nu }\ ^{\mathbf{g}}\mathbf{R}_{_{\mu \nu
}}^{\quad \gamma \tau }\ ^{\mathbf{g}}\mathbf{R}_{_{\gamma \tau }}^{\quad
\gamma \tau },
\end{equation*}%
which may written in equivalent form in terms of the Weyl tensor as it was
obtained by certain simplified computations for the Levi--Civita connection
in \cite{vanven}. Choosing a d--connection $^{\mathbf{g}}\mathbf{\Gamma }_{\
\beta \gamma }^{\alpha }=\widehat{\mathbf{\mathring{\Gamma}}}_{\ \beta
\gamma }^{\alpha }$ we impose the nonholonomic constraints (\ref{constr}),
i.e. vanishing of sub--integral coefficients [with respect to N--adapted
frames, we get constant curvature coefficients with $\ \widehat{\mathbf{%
\mathring{R}}}_{\ \beta ^{\prime }\gamma ^{\prime }\delta ^{\prime
}}^{\alpha ^{\prime }}=0$ (\ref{auxdcurv}) and $\ \widehat{\mathbf{\mathring{%
R}}}_{\ \beta ^{\prime }\gamma ^{\prime }}=0$ (\ref{vacdeinst})] resulting
in $\mathbf{\Gamma }_{\infty }^{(2)}=0.$ This shows that pure gravity may be
two--loop nondivergent, even on shell, but for an alternative d--connection
which is also uniquely constructed from the metric coefficients (we can not
obtain such a result if we work directly only with the Levi--Civita
connection). So, there are hidden symmetries operating on the gravitational
sector and they are related to the possibility that for a metric tensor we
can construct an infinite number of metric compatible d--connections, all
determined by this metric tensor with respect to a prescribed nonholonic
structure of frames. As we shall see in the next subsection, this fact
carries new possibilities to avoid constructions with infinitely many
couplings and higher order curvature terms.

\subsection{Renormalizaton of gravity with infinitely many couplings}

The final goal of a well--defined perturbation theory is the resummation of
the series expansion which has to be performed at least in suitable
correlation functions and physical quantities. For such constructions, the
terms that cannot be reabsorbed by means of field redefinitions have to be
reabsorbed by means of redefinitions of the coupling constants. When the
classical action does not contain the necessary coupling constants $\lambda
, $ new coupling constants have to be introduced. The Einstein gravity
theory is not renormailzable, which means that divergences can be removed
only at the price of introducing infinitely many coupling constants. In Ref. %
\cite{anselmi}, it was shown that the problem of renormalization of
theories of gravity with infinitely many couplings can be solved when the
spacetime manifold admits a metric of constant curvature. It was also proven
that is possible to screen the terms of a generalized gravitational
Lagrangian when, for instance, a whole class of terms is not turned on by
renormalization, if it is absent at the tree level.

Working with d--connections, we can generalize the Anselmi's constructions
for arbitray (pseudo) Riemannian metrics because we can always define such
nonholonomic distributions when a given metric and the curvature of certain
d--connections are characterized by constant matrix coefficients. We can
give a physical sence to such quantum gravity models with infinitely many
parameters and nonholonomic distributions at arbitrary energies if we show
that the formalism does not drive an unitary propagator into a non--unitary
(i.e. higher--derivative) propagator.\footnote{%
One might happen that the non-renormalizability of the Einstein gravity
theory and its nonholonomic deformations can potentially generate all sorts
of counterterms, including those that can affect the propagator with
undesirable higher derivatives.}

A generalized gravitational action (supposed to be more convenient for
purposes of renormalization of gravitational interactions) \textbf{\ }%
contains infinitely many couplings, but not all of the ones that might have
been expected, and in the nonholonomic formalism one can be considered such
constraints when only a finite number of terms are nonzero. In quantum
gravity based on the Levi--Civita connection, the metric of constant
curvature is an extremal, but not a minimum, of the complete action, which
results in the problem how to fix a ''right'' perturbative vacuum. For a
correspondingly defined d--connection, it appears to be possible
to introduce a good perturbative vacuum, choosing such nonholonomic
distributions the curvature is negative and stating the conditions when such
a nonholonomic quantum vacuum has a negative asymptotically constant
curvature. Such properties may not be true for the Levi--Civita connection,
but we can always extract it from a right perturbative nonholonomic vacuum
and N--adapted quantum perturbations of a suitable metric compatible
d--connection.

For a d--connection $^{\mathbf{g}}\mathbf{D}$ with curvature d--tensor $\ ^{%
\mathbf{g}}\mathbf{R}_{\alpha \beta \mu \nu },$ we introduce the values%
\begin{equation*}
\ ^{\mathbf{g}}\mathbf{\check{R}}_{\alpha \beta \mu \nu }=\ ^{\mathbf{g}}%
\mathbf{R}_{\alpha \beta \mu \nu }-\frac{\Lambda }{6}(\mathbf{g}_{\alpha \mu
}\mathbf{g}_{\beta \nu }-\mathbf{g}_{\alpha \nu }\mathbf{g}_{\beta \mu })
\end{equation*}%
and
\begin{equation*}
\ ^{\mathbf{g}}\mathbf{\check{G}=}\ ^{\mathbf{g}}\mathbf{R}_{\alpha \beta
\mu \nu }\ ^{\mathbf{g}}\mathbf{R}^{\alpha \beta \mu \nu }-4\ ^{\mathbf{g}}%
\mathbf{R}_{\alpha \beta }\ ^{\mathbf{g}}\mathbf{R}^{\alpha \beta }+\ ^{%
\mathbf{g}}\overleftrightarrow{\mathbf{R}}^{2},
\end{equation*}%
which are convenient to study $\mathbf{\rho }$--expansions if we choose an
appropriate gravitational vacuum $\underline{\mathbf{g}}_{\alpha \nu }$ to
define quantum fluctuations $\mathbf{\rho }_{\alpha \nu },$ where $\mathbf{g}%
_{\alpha \nu }=\underline{\mathbf{g}}_{\alpha \nu }+\mathbf{\rho }_{\alpha
\nu }.$ By inductive hypothesis, assuming that the $\mathcal{O}(\mathbf{\rho
})$-- and $\mathcal{O}(\mathbf{\rho }^{2})$--contributions come only from
Lagrange density $\ ^{0}\mathcal{L}$ (\ref{ld0}), we introduce a generalized
Lagrange density
\begin{equation}
\mathcal{L}=\frac{1}{\kappa ^{2}}\sqrt{\mathbf{g}}\left[ -\ ^{\mathbf{g}}%
\overleftrightarrow{\mathbf{R}}+\Lambda +\lambda \kappa ^{2}\ ^{\mathbf{g}}%
\mathbf{\check{G}}+\sum\limits_{s=1}^{\infty }\lambda _{s}\kappa ^{2s+2}%
\mathcal{F}_{s}[\ ^{\mathbf{g}}\mathbf{D},\ ^{\mathbf{g}}\mathbf{\check{R}}%
,\Lambda ]\right] ,  \label{ldg}
\end{equation}%
where $\lambda $ and $\lambda _{s}$ label the set of infinite many
couplings. A term $\mathcal{F}_{s}$ in (\ref{ldg}) is a collective
denotation for gauge--invariant terms of dimension $2s+4$ which can be
constructed from three or more curvature d--tensors, $\ ^{\mathbf{g}}\mathbf{%
R}_{\alpha \beta \mu \nu },$ d--connection, \ $\ ^{\mathbf{g}}\mathbf{D,}$
and powers of $\Lambda ,$ in our approach, up to total derivatives adapted
to N--connection structure. It should be noted that index contracted values $%
\ ^{\mathbf{g}}\mathbf{\check{R}}_{\alpha \beta }$ and $\ ^{\mathbf{g}}%
\mathbf{\check{R}}_{\alpha }^{\ \alpha }$ can be removed from $\mathcal{F}%
_{s}$ by means of field redefinitions. In the first approximation, $\mathcal{%
F}_{1}$ contains a linear combination for three multiples $\ ^{\mathbf{g}}%
\mathbf{\check{R}}_{\cdot \cdot \cdot \cdot }\ ^{\mathbf{g}}\mathbf{\check{R}%
}_{\cdot \cdot \cdot \cdot }\ ^{\mathbf{g}}\mathbf{\check{R}}_{\cdot \cdot
\cdot \cdot }$ with all possible contractions of indices, but does not
contain terms like $\ ^{\mathbf{g}}\mathbf{\check{R}}_{\cdot \cdot \cdot
\cdot }\mathbf{\ }^{\mathbf{g}}\mathbf{D\ }^{\mathbf{g}}\mathbf{D}\ ^{%
\mathbf{g}}\mathbf{\check{R}}_{\cdot \cdot \cdot \cdot }$ which would affect
the $\mathbf{\rho } $--propagator with higher derivatives.

In Ref. \cite{anselmi}, for $\ ^{\mathbf{g}}\mathbf{D=\ }^{\mathbf{g}}%
\mathbf{\nabla ,}$ it is proven that the gravity theory derived for
Lagrangian density $\mathcal{L}$ (\ref{ldg}) is renormalizable in the sense
that such a Lagrangian preserves its form under renormalization in arbitrary
spacetime dimension greater than two. Nevertheless, the gravitational field
equations obtained from (\ref{ldg}) are not just the Einstein equations and
the quantum vacuum for this theory must have a negative asymptotically
constant curvature. It was also concluded that if the theory has no
cosmological constant or the space-time manifold admits a metric of constant
curvature, the propagators of the fields are not affected by higher derivatives.

Working with d--connections, we can always impose nonholonomic constraints
when conditions of type (\ref{constr}) are satisfied which results in an
effective nonholonomic gravity model with a finite number of couplings. As a
matter of principle, we can limit our computations only to two--loop
constructions, because terms $\mathcal{F}_{s}$ can be transformed in zero
for a corresponding nonholonomic distribution, with a formal renormalization
of such a model. Of course, being well defined as a perturbative quantum
model such a classical theory for $\ ^{\mathbf{g}}\mathbf{D}$ is not
equivalent to Einstein gravity. Nevertheless, we can always add the
contributions of distortion tensor quantized as a nonholonomic gauge gauge
theory and reconstruct the classical Einstein theory and its perturbative
quantum corrections renormalized both by nonholonomic geometric methods
combined with standard methods elaborated for Yang--Mills fields.

\section{Nonholonomic Differential Renormalization}

In this section we shall extend the method of Differential Renormalization
(DiffR) \cite{freed} on N--anholonomic backgrounds, which in our approach is
to be elaborated as a renormalization method in real space when too singular
coordinate--space expressions are replaced by N--elongated partial
derivatives of some corresponding less singular values. In brief, we shall
denote this method Diff$_{N}$R. It should be noted here that differential
renormalization keep all constructions in four dimension which is not the
case for dimensional regularization or dimensional reductions.

Our goal is to formulate a renormalization procedure for a nonholonomic de
Sitter frame gauge gravity theory with field equations (\ref{efymeq1}) when
gravitational distortion $\ _{\shortmid }\widehat{\mathbf{\mathring{Z}}}_{\
\beta \gamma }^{\alpha }$ is encoded into geometric structures on total
space.\footnote{%
for simplicity, we shall quantize a model with zero matter field source} The
base spacetime nonholonomic manifold $\mathbf{V}$ is considered to be
endowed with a d--connection structure $\ ^{\circ }\widehat{\mathbf{D}}=\{%
\widehat{\mathbf{\mathring{\Gamma}}}_{\ \beta \gamma }^{\alpha }\}$
determined completely by a metric, $\mathbf{g=\mathring{g},}$ $\ $\ and
N--connection, $\mathring{N}_{i^{\prime }}^{a^{\prime }},$ structures. The
linear connection $\ ^{\circ }\widehat{\mathbf{D}}$ is with constant
curvature coefficients subjected to the conditions $\ \widehat{\mathbf{%
\mathring{R}}}_{\ \beta ^{\prime }\gamma ^{\prime }}=\ \widehat{\mathbf{%
\mathring{R}}}_{\ \beta ^{\prime }\gamma ^{\prime }\alpha ^{\prime
}}^{\alpha ^{\prime }}=0$ (\ref{vacdeinst}) and (\ref{constr}) allowing us
to perform formal one-- and two--loop renormalization of the nonholonomic
background $\mathbf{V}$ as we discussed in previous section \ref{stlc}.

\subsection{Two--loop diagrams on nonholonomic backgrounds and Diff$_{N}$R}
We sketch some key constructions how differential renormalization formalism and loop diagrams can be generalized on nonholonomic spaces.
\subsubsection{Propagators on N--anholonomic manifolds}
Let us consider the massless propagator%
\begin{equation*}
\bigtriangleup (\ ^{1}u-\ ^{2}u)\equiv \ ^{12}\bigtriangleup =\ _{\
^{2}u}^{\ ^{1}u}\bigtriangleup =\frac{1}{(2\pi )^{2}}\frac{1}{(\ ^{1}u-\
^{2}u)^{2}},
\end{equation*}%
for two points $\ ^{1,2}u=(\ ^{1,2}x^{i},\ ^{1,2}y^{a})\in \mathbf{V.}$ On a
pseudo--Euclidean spacetime, this propagator defines the one--loop
contribution of so--called scalar $\lambda \phi ^{4}(u)$ theory
\begin{eqnarray*}
\Gamma (\ ^{1}u,\ ^{2}u,\ ^{3}u,\ ^{4}u) &=&\frac{\lambda ^{2}}{2}[\
^{(4)}\delta (\ ^{1}u-\ ^{2}u)\ ^{(4)}\delta (\ ^{3}u-\ ^{4}u)\left[
\bigtriangleup (\ ^{1}u-\ ^{4}u)\right] ^{2} \\
&&+(\mbox{2 points permutations})],
\end{eqnarray*}%
where $^{(4)}\delta =\delta $ is the four dimensional delta function. The
usual DiffR method proposes to replace the function $\frac{1}{u^{4}}$ (which
does not have a well defined Fourier transform), for $u\neq 0,$ with the
Green function $G(u^{2})$, i.e. solution of
\begin{equation*}
\frac{1}{u^{4}}=\square G(u^{2}),
\end{equation*}%
for d'Alambertian $\square =\partial ^{\mu }\partial _{\mu }$ determined by
partial derivatives $\partial _{\mu }$ and metric on Minkowski space. This
solution (renormalized, with left label $R$) is
\begin{equation*}
\frac{1}{u^{4}}\rightarrow \ ^{R}\left\lfloor \frac{1}{u^{4}}\right\rfloor =-%
\frac{1}{4}\square \frac{\ln u^{2}\varpi ^{2}}{u^{2}}
\end{equation*}%
where the constant $\varpi $ of mass dimension is introduced for dimensional
reasons. This constant parametrizes a local ambiguity
\begin{equation*}
\square \frac{\ln u^{2}\varpi ^{\prime 2}}{u^{2}}=\square \frac{\ln
u^{2}\varpi ^{2}}{u^{2}}+2\ln \frac{\varpi ^{\prime }}{\varpi }\delta (u),
\end{equation*}%
when the shift $\varpi \rightarrow \varpi ^{\prime }$ can be absorbed by
rescaling the constant $\lambda ,$ see details in Ref. \cite{freed}; we can
related this property with the fact that renormalized amplitudes are
constrained to satisfy certain renormalization group equations with $\varpi $
being the renormalization group scale. It should be noted here that both
non--renormalized and renormalized expressions coincide for $u\neq 0,$ but
that with label ''R'' has a well defined fourier transform (if we neglect
the divergent surface terms that appears upon integrating by parts). For
instance, we have
\begin{eqnarray*}
\int \ ^{R}\left\lfloor \frac{1}{u^{4}}\right\rfloor e^{ip\cdot u}d^{4}u &=&-%
\frac{1}{4}\int \square \frac{\ln u^{2}\varpi ^{2}}{u^{2}}e^{ip\cdot
u}d^{4}u= \\
&=&\frac{p^{2}}{4}\int \frac{\ln u^{2}\varpi ^{2}}{u^{2}}e^{ip\cdot
u}d^{4}u=-\pi ^{2}\ln \frac{p^{2}}{\varpi ^{2}}.
\end{eqnarray*}

On a N--anholonomic background $\mathbf{V}$ enabled with constant
coefficients $\mathring{g}_{\alpha \beta }$ and $\mathring{N}_{i^{\prime
}}^{a^{\prime }},$ see formulas (\ref{ddifc}) and (\ref{rediff}), the
d'Alambert operator $\mathring{\square}=\ ^{\circ }\widehat{\mathbf{D}}^{\mu
^{\prime }}$ $\ ^{\circ }\widehat{\mathbf{D}}_{\mu ^{\prime }},$ where $%
^{\circ }\widehat{\mathbf{D}}_{\mu ^{\prime }}=\mathbf{\mathring{e}}_{\mu
^{\prime }}\pm \widehat{\mathbf{\mathring{\Gamma}}}_{\ \mu ^{\prime }\cdot
}^{\cdot }$ contains the N--elongated partial derivative $\mathbf{\mathring{e%
}}_{\mu ^{\prime }}=(\mathbf{\mathring{e}}_{i^{\prime }}=\partial
_{i^{\prime }}-\mathring{N}_{i^{\prime }}^{a^{\prime }}(u)\partial
_{a^{\prime }},\partial _{b^{\prime }})$ as the dual to $\mathbf{\mathring{e}%
}^{\nu ^{\prime }}.$ We suppose that the nonholonomic structure on $\mathbf{V%
}$ is such way prescribed that we have a well defined background operator $%
\mathring{\square}$ constructed as a quasi--linear combination (with
coefficients depending on $u^{\alpha }$) of partial derivatives $\partial
_{\mu }.$ In the infinitesimal vicinity of a point $\ ^{0}u^{\alpha },$ we
can always consider $\mathring{\square}$ to be a linear transform (depending
on values $\mathring{g}_{\alpha \beta }$ and $\mathring{N}_{i^{\prime
}}^{a^{\prime }}$ in this point) of the flat operator $\square .$
Symbolically, we shall write
\begin{equation*}
\int \ ^{\mathring{R}}\left\lfloor \frac{1}{u^{4}}\right\rfloor e^{ip\cdot u}%
\sqrt{|\mathring{g}_{\alpha \beta }|}d^{4}u=-\pi ^{2}\mathring{l}n\frac{p^{2}%
}{\varpi ^{2}}
\end{equation*}%
for the formal solution of
\begin{equation*}
\frac{1}{u^{4}}=\mathring{\square}G(u^{2})
\end{equation*}%
with
\begin{equation*}
\frac{1}{u^{4}}\rightarrow \ ^{\mathring{R}}\left\lfloor \frac{1}{u^{4}}%
\right\rfloor =-\frac{1}{4}\mathring{\square}\frac{\mathring{l}n\left(
u^{2}\varpi ^{2}\right) }{u^{2}}.
\end{equation*}%
Such functions can be computed in explicit form, as certain series, for a
prescribed nonhlononomic background, for instance, beginning with a
classical exact solution of the Einstein equations and a fixed 2+2
splitting. For such computations, we use formal integration by parts and
have to consider a ''locally anisotropic'' ball $\mathcal{B}_{\varepsilon }$
of radius $\varepsilon $ around a point $\ ^{0}u\in \mathbf{V}$ and keep
surface terms (we denote such an infinitesimal term by $\mathring{\delta}%
\sigma ^{\mu ^{\prime }}$ and a closed region $\mathcal{S}_{\varepsilon }$
in $\mathbf{V}$) like in formulas related to integration of a function $%
A(u), $ with volume element $dV(u)=\sqrt{|\mathring{g}_{\alpha \beta }|}%
d^{4}u,$%
\begin{eqnarray*}
\int A(u)\frac{1}{u^{4}}d\ ^{u}V &=&-\frac{1}{4}\int A(u)\left( \mathring{%
\square}\frac{\mathring{l}n\left( u^{2}\varpi ^{2}\right) }{u^{2}}\right)
dV(u) \\
&=&-\frac{1}{4}\int \left( \mathring{\square}A(u)\right) \frac{\mathring{l}%
n\left( u^{2}\varpi ^{2}\right) }{u^{2}}dV(u)
\end{eqnarray*}%
and%
\begin{eqnarray*}
&&\int_{\mathbf{V/}\mathcal{B}_{\varepsilon }}A(u)\left( \mathring{\square}%
\frac{\mathring{l}n\left( u^{2}\varpi ^{2}\right) }{u^{2}}\right)
dV(u)=\int_{\mathcal{S}_{\varepsilon }}A(u)\ ^{\circ }\widehat{\mathbf{D}}%
_{\mu ^{\prime }}\frac{\mathring{l}n\left( u^{2}\varpi ^{2}\right) }{u^{2}}%
\mathring{\delta}\sigma ^{\mu ^{\prime }} \\
&&-\int_{\mathbf{V/}\mathcal{B}_{\varepsilon }}\left( \ ^{\circ }\widehat{%
\mathbf{D}}_{\mu ^{\prime }}A(u\right) \left( \ ^{\circ }\widehat{\mathbf{D}}%
^{\mu ^{\prime }}\frac{\mathring{l}n\left( u^{2}\varpi ^{2}\right) }{u^{2}}%
\right) dV(u).
\end{eqnarray*}

We can approximate the first integral in the last formula as
\begin{equation*}
\int_{\mathcal{S}_{\varepsilon }}A(u)\left( \ ^{\circ }\widehat{\mathbf{D}}%
_{\mu ^{\prime }}\frac{\mathring{l}n\left( u^{2}\varpi ^{2}\right) }{u^{2}}%
\right) \mathring{\delta}\sigma ^{\mu ^{\prime }}\rightarrow 4\pi ^{2}A(\
^{0}u)(1-\ln \varepsilon ^{2}\varpi ^{2})+\mathcal{O}(\varepsilon )
\end{equation*}%
which is divergent for $\varepsilon \rightarrow 0.$ So, we have a formal
integration rule by parts because we use conterterms. Nevertheless, this
method of regularization does not require an explicit use of conterterms in
calculations even the background space may be subjected to nonholonomic
constraints.

\subsubsection{Higher loops}

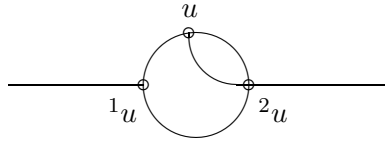
\begin{figure}[tbph]
\begin{center}
\setlength{\unitlength}{1cm}
\begin{picture}(4,3)
\thinlines
\put(2.5,1){\circle{7}}
\put(0,1){\line(1,0){1.8}}
\put(3.2,1){\line(1,0){1.8}}
\put(1.2,0.5){$\ ^1u$}
\put(3.2,0.5){$\ ^2u$}
\put(2.3,1.9){$u$}
\put(3.5,1.7){\oval(2.2,1.4)[bl]}
\put(2.3,1.6){$\circ$}
\put(1.7,0.9){$\circ$}
\put(3.1,0.9){$\circ$}
\end{picture}
\end{center}
\caption{\textbf{A two--loop diagram with nestled divergences}}
\label{fig1}
\end{figure}

The method Diff$_{N}$R can be applied also to multi--loop expressions. As an
example, we consider a two loop diagram from Figure \ref{fig1}, $%
\bigtriangleup (\ ^{1}u-\ ^{2}u)\ ^{1}I(\ ^{1}u-\ ^{2}u),$ for
\begin{equation}
\ ^{1}I(\ ^{1}u-\ ^{2}u)=\int \ \bigtriangleup (\ ^{1}u-u)\left(
\bigtriangleup (u-\ ^{2}u)\right) ^{2}dV(u).  \label{2ld}
\end{equation}%
We get divergences whenever two points come together. Proceeding recursively
(starting from the most inner divergence), we can renormalize them,%
\begin{eqnarray*}
\ ^{\mathring{R}}\left\lfloor \bigtriangleup (\ ^{1}u-\ ^{2}u)\int \
\bigtriangleup (\ ^{1}u-u)\ \ ^{\mathring{R}}\left\lfloor \left(
\bigtriangleup (u-\ ^{2}u)\right) ^{2}\right\rfloor \ dV(u)\right\rfloor &=&
\\
\ ^{\mathring{R}}\left\lfloor -\frac{1}{4(2\pi )^{8}}\frac{1}{(\ ^{1}u-\
^{2}u)^{2}}\int \frac{1}{(\ ^{1}u-u)^{2}}\mathring{\square}\frac{\mathring{l}%
n\left[ (\ ^{2}u-u)^{2}\ ^{1}\varpi ^{2}\right] }{(\ ^{2}u-u)^{2}}\
dV(u)\right\rfloor \ && \\
=\ ^{\mathring{R}}\left\lfloor -\frac{1}{4(2\pi )^{6}}\frac{\mathring{l}n%
\left[ (\ ^{2}u-u)^{2}\ ^{1}\varpi ^{2}\right] }{(\ ^{2}u-u)^{2}}\
\right\rfloor && \\
=-\frac{1}{32(2\pi )^{6}}\mathring{\square}\frac{\left( \mathring{l}n\left[
(\ ^{1}u-\ ^{2}u)^{2}\ ^{1}\varpi ^{2}\right] \right) ^{2}+2\ \mathring{l}n%
\left[ (\ ^{1}u-\ ^{2}u)^{2}\ ^{2}\varpi ^{2}\right] }{(\ ^{1}u-\ ^{2}u)^{2}}%
, &&
\end{eqnarray*}%
where there are considered two constants $\ ^{1}\varpi ^{2}$ and $\
^{2}\varpi ^{2}$ and integrating by parts the ''anisotropic'' d'Alambertian
we used the local limit $\mathring{\square}\rightarrow \square ,$ when $%
\square (\ ^{1}u-\ ^{2}u)^{-2}=\delta (\ ^{1}u-\ ^{2}u).$

Using nonholonomic versions of d'Alambertian and logarithm function, i.e. $%
\mathring{\square}$ and $\mathring{l}n,$ corresponding N--adapted partial
derivative operators and differentials, $\mathbf{\mathring{e}}_{\mu ^{\prime
}}$ and $\mathbf{\mathring{e}}^{\nu ^{\prime }},$ and their covariant
generalizations with $\ ^{\circ }\widehat{\mathbf{D}}_{\mu ^{\prime }},$ we
can elaborate a systematic N--adapted differential renormalization procedure
to all orders in pertrurbations theory, extending the constructions from
Ref. \cite{latorre}. This procedure maintains unitarity, fulfills locality
and Lorentz invariance to all orders and allows to renormalize massive
fields (in our nonholonomic gauge like approach to Einstein gravity, masses
have to be considered for quantum systems of gravitational and matter field
equations; for simplicity, we omit such constructions in this paper which
are similar to holonomic ones for matter and usual Yang--Mills fields in %
\cite{freed}). Such nonholonomic implementations of Bogoliubov's $R$%
--operator (this operation yields directly renormalized correlation
functions satisfying renormalization group equations) in momentum spaces can
be also applied to expression with IR divergences, when $p^{\mu ^{\prime
}}\rightarrow 0$ and UV divergences, when $p^{\mu ^{\prime }}\rightarrow
\infty .$ The corresponding recursion formulas from\ Refs. \cite%
{seijas,avd,smirnov} can be easily re--defined for nonholonomic backgrounds
with constant $\mathring{g}_{\alpha \beta }$ and correspondingly defined
connections $\mathring{N}_{i^{\prime }}^{a^{\prime }}$ and $\ ^{\circ }%
\widehat{\mathbf{D}}_{\mu ^{\prime }}.$

\subsection{Constrained differential renormalization on nonholo\-no\-mic
spac\-es}

We can apply the method of constrained differential renormalization, see a
review and basic references in \cite{aguila} in order to avoid the necessity
of imposing Ward identities in each calculation scheme. The constructions
(in brief, we shall write for this method CDR$_{N}$) \ can be adapted to the
N--anholonomic structure as we have done in the previous section. One should
follow the rules:

\begin{enumerate}
\item N--adapted differential reduction;

\begin{description}
\item One reduces to covariant d--derivatives of logarithmically divergent
(at most), without introducing extra dimensional constants, all functions
with singularities worse than logarithmic ones.

\item One introduces a constant $\varpi $ (it has dimension of mass and
plays the role of renormalization group scale) for any logarithmically
divergent expression which allows us to rewrite such an expression as
derivatives of regular functions.

\item In infinitesimal limits, the ''anisotropic'' logarithm $\mathring{l}n$
and operator $\mathring{\square}$ can transform into usual ones in flat
spacetimes.
\end{description}

\item Integration by parts using N--elongated differentials $\mathbf{%
\mathring{e}}^{\nu ^{\prime }}$(\ref{ddifc}). It is possible to omit
consideration of divergent surface terms that appear under integration by
parts. For N--adapted differentiation and renormalization of an arbitrary
function $A(u),$ we have $\ ^{\mathring{R}}\left\lfloor \ ^{\circ }\widehat{%
\mathbf{D}}A\right\rfloor =\ ^{\circ }\widehat{\mathbf{D}}\ ^{\mathring{R}%
}\left\lfloor A\right\rfloor $ and $\ ^{\mathring{R}}\left\lfloor \mathbf{%
\mathring{e}}A\right\rfloor =\mathbf{\mathring{e}}\ ^{\mathring{R}%
}\left\lfloor A\right\rfloor .$

\item Renormalization of delta function and propagator equation,%
\begin{eqnarray*}
\ ^{\mathring{R}}\left\lfloor A(u,\ ^{1}u,\ ^{2}u,...,\ ^{k-1}u)\ \delta (\
^{k}u-u)\right\rfloor &=& \\
\ \ ^{\mathring{R}}\left\lfloor A(u,\ ^{1}u,\ ^{2}u,...,\
^{k-1}u)\right\rfloor \delta (\ ^{k}u-u), && \\
&& \\
\ ^{\mathring{R}}\left\lfloor A(u,\ ^{1}u,\ ^{2}u,...,\ ^{k}u)\ \left(
\mathring{\square}-\mu ^{2}\right) \ ^{\mu }\bigtriangleup (\
^{k}u-u)\right\rfloor &=& \\
-\ \ ^{\mathring{R}}\left\lfloor A(u,\ ^{1}u,\ ^{2}u,...,\ ^{k-1}u)\ \delta
(u)\right\rfloor , &&
\end{eqnarray*}%
where $^{\mu }\bigtriangleup $ is the propagator of a particle of mass $\mu
, $ where $\mu =0$ for gravitational fields, \ and $A$ is an arbitrary
function.
\end{enumerate}

The method CDR$_{N}$ contains two steps:

\begin{itemize}
\item The Feynman diagrams are expressed in terms of basic functions
performing all index contractions (this method does not commute with
contractions of indices); using the Leibniz rule, we move all N--adapted
derivatives to act on one of the propagators.

\item Finally, we replace the basic functions with their renormalized
versions.
\end{itemize}

In order to understand how the above mentioned method should be applied in
explicit computations, we present a series of important examples.

The one--loop correction to the two--point function in $\lambda \phi ^{4}(u)$
theory is defined by renormalization of $\bigtriangleup (u)\delta (u),$
which is constrained by above mentioned rules to result in $\ \ ^{\mathring{R%
}}\left\lfloor \bigtriangleup (u)\delta (u)\right\rfloor =0.$ This way we
get that all massless one--point functions in CDR$_{N}$ are zero, i.e. a
nonholonomic structure does not change similar holonomic values.

A nonholonomic configuration can be included into an operator containing a
N--adapted covariant derivative, but also results in a zero constribution if
the operator $\mathring{\square}$ is introduced into consideration, i.e. $\
^{\mathring{R}}\left\lfloor \bigtriangleup \mathring{\square}\bigtriangleup
\right\rfloor (u)=0.$ One hold true the important formulas:%
\begin{eqnarray*}
\ \ ^{\mathring{R}}\left\lfloor \bigtriangleup ^{2}\right\rfloor (u) &=&-%
\frac{1}{4(2\pi )^{4}}\mathring{\square}\frac{\mathring{l}n\left[ u^{2}\
\varpi ^{2}\right] }{u^{2}}, \\
\ \ ^{\mathring{R}}\left\lfloor \bigtriangleup \ ^{\circ }\widehat{\mathbf{D}%
}_{\mu ^{\prime }}\bigtriangleup \right\rfloor (u) &=&-\frac{1}{8(2\pi )^{4}}%
\ ^{\circ }\widehat{\mathbf{D}}_{\mu ^{\prime }}\left( \mathring{\square}%
\frac{\mathring{l}n\left[ u^{2}\ \varpi ^{2}\right] }{u^{2}}\right) , \\
\ \ ^{\mathring{R}}\left\lfloor \bigtriangleup \ ^{\circ }\widehat{\mathbf{D}%
}_{\mu ^{\prime }}\bigtriangleup \right\rfloor (u) &=&-\frac{1}{12(2\pi )^{4}%
}\ (\ ^{\circ }\widehat{\mathbf{D}}_{\mu ^{\prime }}\ ^{\circ }%
\widehat{\mathbf{D}}_{\nu ^{\prime }}-\frac{\delta _{\mu ^{\prime }\nu
^{\prime }}}{4}\mathring{\square}) \left( \mathring{\square}\frac{%
\mathring{l}n\left[ u^{2}\ \varpi ^{2}\right] }{u^{2}}\right) \\
&&+\frac{1}{288\pi ^{2}} (\ ^{\circ }\widehat{\mathbf{D}}_{\mu ^{\prime
}}\ ^{\circ }\widehat{\mathbf{D}}_{\nu ^{\prime }}-\delta _{\mu ^{\prime
}\nu ^{\prime }}\mathring{\square}) \delta (u).
\end{eqnarray*}

The method CDR$_{N}$ can be applied to more than two propagators. For
instance, we can write $T[\mathcal{O}]=\bigtriangleup \bigtriangleup
\mathcal{O}\bigtriangleup ,$ for three propagators, and compute%
\begin{eqnarray*}
\ \ ^{\mathring{R}}T\left[ \ ^{\circ }\widehat{\mathbf{D}}_{\mu ^{\prime }}\
^{\circ }\widehat{\mathbf{D}}_{\nu ^{\prime }}\right] &=&\ \ ^{\mathring{R}}T%
\left[ \ ^{\circ }\widehat{\mathbf{D}}_{\mu ^{\prime }}\ ^{\circ }\widehat{%
\mathbf{D}}_{\nu ^{\prime }}-\frac{\delta _{\mu ^{\prime }\nu ^{\prime }}}{4}%
\mathring{\square}\right] +\frac{\delta _{\mu ^{\prime }\nu ^{\prime }}}{4}%
\left[ \mathring{\square}\right] \\
&&-\frac{1}{128\pi ^{2}}\delta _{\mu ^{\prime }\nu ^{\prime }}\delta (\
^{1}u)\delta (\ ^{2}u),
\end{eqnarray*}%
for two points $\ ^{1}u$ and $\ ^{2}u.$

\subsection{Using one--loop results for CDR$_{N}$ in two--loop calculus}

The CDR$_{N}$ can be easily developed at loop--order higher than one, which
is enough to define the renormalization group (RG) equations. We restrict
our geometric analysis only for such constructions and do not analyze, for
instance, scattering amplitudes.

For the simplest, so--called nested divergences, we can compute in
N--adapted form (applying formulas from previous section) the value (\ref%
{2ld}), when according the CDR$_{N}$ $\ $rules, we get the renormalized
values
\begin{eqnarray*}
\ \ ^{\mathring{R}}\left\lfloor \ ^{1}I\right\rfloor (\ u) &=&\frac{1}{%
4(2\pi )^{4}}\frac{\mathring{l}n\left[ u^{2}\ \varpi ^{2}\right] }{u^{2}}%
+..., \\
\ \ ^{\mathring{R}}\left\lfloor \bigtriangleup \ ^{1}I\right\rfloor (\ u)
&=&-\frac{1}{32(2\pi )^{6}}\mathring{\square}\frac{\left( \mathring{l}n\left[
u^{2}\ \varpi ^{2}\right] \right) ^{2}+2\mathring{l}n\left[ u^{2}\ \varpi
^{2}\right] }{u^{2}}+..., \\
\ ^{\mathring{R}}\left\lfloor \bigtriangleup \ ^{\circ }\widehat{\mathbf{D}}%
_{\mu ^{\prime }}\ ^{1}I\right\rfloor (\ u) &=&-\frac{1}{96(2\pi )^{6}}[%
\mathring{\square}\frac{\left( \mathring{l}n\left[ u^{2}\ \varpi ^{2}\right]
\right) ^{2}+2\mathring{l}n\left[ u^{2}\ \varpi ^{2}\right] }{u^{2}}- \\
&&-\frac{\delta _{\mu ^{\prime }\nu ^{\prime }}}{4}\mathring{\square}^{2}%
\frac{\left( \mathring{l}n\left[ u^{2}\ \varpi ^{2}\right] \right) ^{2}+%
\frac{11}{3}\mathring{l}n\left[ u^{2}\ \varpi ^{2}\right] }{u^{2}}]+..., \\
\ \ ^{\mathring{R}}\left\lfloor \bigtriangleup \mathring{\square}\
^{1}I\right\rfloor (\ u) &=&\frac{1}{32(2\pi )^{4}}\mathring{\square}^{2}%
\frac{\mathring{l}n\left[ u^{2}\ \varpi ^{2}\right] }{u^{2}}+....,
\end{eqnarray*}%
where ''...'' stand for the two--loop local terms that are not taken into
account.

In order to compute overlapping divergences, we define for any differential
d--operator $\ ^{i}\mathcal{O},$ and (for instance) $\ _{\ ^{1}u}^{i}%
\mathcal{O}$ taken in the point $\ ^{1}u,$ the value $H(\ ^{1}u-\
^{2}u)\equiv H(u),$ for $\ ^{\circ }\widehat{\mathbf{D}}_{\mu ^{\prime }}$
taken in the point $\ ^{1}u,$%
\begin{eqnarray*}
H[\ ^{1}\mathcal{O},\ ^{2}\mathcal{O};\ ^{3}\mathcal{O},\ ^{4}\mathcal{O}]
&=&\int \ _{\ v}^{\ u}\bigtriangleup \left( \ _{\ ^{1}u}^{1}\mathcal{O}\ _{\
u}^{\ ^{1}u}\bigtriangleup \right) \left( \ _{\ ^{1}u}^{2}\mathcal{O}\ _{\
v}^{\ ^{1}u}\bigtriangleup \right) \\
&&\left( \ _{\ ^{2}u}^{3}\mathcal{O}\ _{\ u}^{\ ^{2}u}\bigtriangleup \right)
\left( \ _{\ ^{2}u}^{4}\mathcal{O}\ _{\ v}^{\ ^{2}u}\bigtriangleup \right)
dV(u)dV(v).
\end{eqnarray*}%
This value is very useful because it can be used as a basis for expressing
the renormalized overlapping contributions to two--point functions in
theories with derivative couplings at two loops. They are necessary if we
need to obtain the beta function following the background field method. The
typical expressions for such renormalized overlapping divergences are
presented in Appendix \ref{asovd}.

\section{Quantization of Distortion Gauge Fields}

\label{sqdgg}This section focuses on quantization of the nonholonomic gauge
model of gravity constructed as a lift of the Einstein theory in the total
space of de Sitter frame bundle. The details of geometric formulation and
classical field equations are given in sections 4.2 and 4.3 of Ref. \cite%
{vpartn1}.

\subsection{Nonholonomic gauge gravity theory}

Let \ $\ ^{\eta }S=\ SO(5)$ be the continuous symmetry/gauge group (in this
model, the isometry group of a de Sitter space $\ ^{5}\Sigma )$ \ with
generators $I^{\underline{1}},...,I^{\underline{S}}$ and structure constants
$f_{\underline{T}}^{\ \ \underline{S}\underline{P}}$ defining a Lie algebra $%
\mathcal{A}_{I}=$ $\mathit{so}(5)$ trough commutation relation
\begin{equation*}
\lbrack I^{\underline{S}},I^{\underline{P}}]=if^{\ \underline{T}\underline{S}%
\underline{P}}I^{\underline{T}},
\end{equation*}%
with summation on repeating indices. The space $\ ^{5}\Sigma $ can be
defined as a hypersurface $\eta _{AB}u^{A}u^{B}=-1$ in a four--dimensional
flat space endowed with a diagonal metric $\eta _{AB}=diag[\pm 1,...,\pm 1],$
where $\{u^{A}\}$ are global Cartezian coordinates in $\mathbb{R}^{5},$
indices $A,B,C...$ run values $1,2,...,5$ and $l>0$ is the constant
curvature of de Sitter space.\footnote{%
A canonical $4+1$ splitting is parametrized by $A=(\underline{\alpha },5),B=(%
\underline{\beta },5),...;\eta _{AB}=(\eta _{\underline{\alpha }\underline{%
\beta }},\eta _{55})$ and $P_{\underline{\alpha }}=l^{-1}M_{5\underline{%
\alpha }},$for $\underline{\alpha },\underline{\beta },...=1,2,3,4$ when the
commutation relations are written%
\begin{eqnarray*}
\lbrack M_{\underline{\alpha }\underline{\beta }},M_{\underline{\gamma }%
\underline{\delta }}] &=&\eta _{\underline{\alpha }\underline{\gamma }}M_{%
\underline{\beta }\underline{\delta }}-\eta _{\underline{\beta }\underline{%
\gamma }}M_{\underline{\alpha }\underline{\delta }}+\eta _{\underline{\beta }%
\underline{\delta }}M_{\underline{\alpha }\underline{\gamma }}-\eta _{%
\underline{\alpha }\underline{\delta }}M_{\underline{\beta }\underline{%
\gamma }}, \\
\lbrack P_{\underline{\alpha }},P_{\underline{\beta }}] &=&-l^{-2}M_{%
\underline{\alpha }\underline{\beta }},\ [P_{\underline{\alpha }},M_{%
\underline{\beta }\underline{\gamma }}]=\eta _{\underline{\alpha }\underline{%
\beta }}P_{\underline{\gamma }}-\eta _{\underline{\alpha }\underline{\gamma }%
}P_{\underline{\beta }}.
\end{eqnarray*}%
This defines a direct sum $\ \mathit{so}(5)=\mathit{so}(4)\oplus \ ^{4}V,$
where $\ ^{4}V$ is the four dimensional vector space stretched on vectors $%
P_{\underline{\alpha }}.$ We remark that $\ ^{4}\Sigma =\ ^{\eta }S/\ ^{\eta
}L,$ where $\ ^{\eta }L=\ SO(4).$ Choosing signature $\eta
_{AC}=diag[-1,1,1,1,1]$ and $\ ^{\eta }S=SO(1,4),$ we get the group of
Lorentz rotations $\ ^{\eta }L=SO(1,3).$} \ In quantum models, it is
convenient to chose the so--called adjoint reprezentation when the
representation matrices are given by the structure constants $\left( \tau ^{%
\underline{S}}\right) _{\underline{T}\underline{P}}=if^{\ \underline{T}%
\underline{S}\underline{P}}.$ Such matrices satisfy the conditions
\begin{equation*}
\sum_{\underline{S}}\left( \tau ^{\underline{S}}\right) ^{2}=\sum_{%
\underline{S}}\tau ^{\underline{S}}\tau ^{\underline{S}}=\ ^{1}C\ \mathbb{I}%
\mbox{\ and \ }tr[\tau ^{\underline{S}}\tau ^{\underline{P}}]=\ ^{2}C\
\delta ^{\underline{S}\underline{P}},
\end{equation*}%
where the quadratic Casimir operator is defined by constant $\ \ ^{1}C,\
^{2}C=const,tr$ denotes trace of matrices and $\mathbb{I}$ is the unity
matrix. In the adjoint representation, we can write $f^{\ \underline{T}%
\underline{S}\underline{P}}f^{\ \underline{L}\underline{S}\underline{P}}=\
^{1}C\delta ^{\underline{T}\underline{L}}.$

With respect to a N--adapted dual basis $\mathbf{\mathring{e}}^{\nu ^{\prime
}}$(\ref{ddifc}) on $\mathbf{V},$ we consider a d--connection (nonholonomic
gauge potential) $\ \mathbf{\mathring{Z}}_{\nu ^{\prime }}=\mathbf{\mathring{%
Z}}_{\nu ^{\prime }}^{\underline{S}}\tau ^{\underline{S}}$ defining the
covariant derivation in the total space $\ ^{\circ }\mathcal{D}_{\nu
^{\prime }}=\ ^{\circ }\widehat{\mathbf{D}}_{\nu ^{\prime }}-i\varkappa
\mathbf{\mathring{Z}}_{\nu ^{\prime }},$ where the constant $\varkappa $ is
an arbitrary one, similar to a particular fixing of d--tensors $\mathbf{Y}%
_{ej}^{m\,},\mathbf{Y}_{mc}^{k\,},\mathbf{Y}_{ck}^{d\,}$ and $\mathbf{Y}%
_{ec}^{d\,}$ in (\ref{mcdc}) in order to state in explicit form a
nonholonomic configuration. If we chose the adjoint representation, we get a
covariant derivation
\begin{equation*}
\ ^{\circ }\mathcal{D}_{\nu ^{\prime }}^{\underline{T}\underline{S}}=\
^{\circ }\widehat{\mathbf{D}}_{\nu ^{\prime }}\delta ^{\underline{T}%
\underline{S}}+\varkappa f^{\ \underline{T}\underline{S}\underline{P}}%
\mathbf{\mathring{Z}}_{\nu ^{\prime }}^{\underline{P}}.
\end{equation*}%
The d--field $\mathbf{\mathring{Z}}_{\nu ^{\prime }}^{\underline{S}}$ is
parametrized in matrix form as a d--connection of type d--connection (\ref%
{conds}) subjected to certain nonholonomic nonlinear gauge and frame
transforms. The curvature is defined by commutator of $\ ^{\circ }\mathcal{D}%
_{\nu ^{\prime }}^{\underline{T}\underline{S}},$
\begin{equation*}
-i\ \ _{\shortmid }\overline{\mathcal{\mathring{Z}}}_{\ \mu ^{\prime }\nu
^{\prime }}^{\underline{P}}\tau ^{\underline{P}}=\left[ \ ^{\circ }\mathcal{D%
}_{\mu ^{\prime }},\ ^{\circ }\mathcal{D}_{\nu ^{\prime }}\right] ,
\end{equation*}
equivalently to $\ _{\shortmid }\overline{\mathcal{\mathring{Z}}}$ from
gauge gravity equations (\ref{efymeq1}). In N--adapted components and
adjoint representation, we get the field strength%
\begin{equation*}
\ \ _{\shortmid }\overline{\mathcal{\mathring{Z}}}_{\ \mu ^{\prime }\nu
^{\prime }}^{\underline{P}}=\ ^{\circ }\widehat{\mathbf{D}}_{\mu ^{\prime }}%
\mathbf{\mathring{Z}}_{\nu ^{\prime }}^{\underline{P}}-\ ^{\circ }\widehat{%
\mathbf{D}}_{\nu ^{\prime }}\mathbf{\mathring{Z}}_{\mu ^{\prime }}^{%
\underline{P}}+\varkappa f^{\ \underline{P}\underline{S}\underline{T}}%
\mathbf{\mathring{Z}}_{\nu ^{\prime }}^{\underline{S}}\mathbf{\mathring{Z}}%
_{\nu ^{\prime }}^{\underline{T}}.
\end{equation*}

For zero source of matter fields, when $\ _{\shortmid }\overline{\mathcal{%
\mathring{J}}}=0$\ (this condition can be satisfied in explicit form for
nonholonomic configurations with $\overline{\mathbf{\mathring{\Gamma}}}=0),$
we can consider a nonholonomic gauge gravity field action (for distortions
of the Levi--Civita connection lifted on total space)
\begin{equation}
\ _{\shortmid }\mathcal{\mathring{S}}(\mathbf{\mathring{Z}})=\frac{1}{4}\int
\ \ _{\shortmid }\overline{\mathcal{\mathring{Z}}}_{\ \ \mu ^{\prime }\nu
^{\prime }}^{\underline{P}}\ \ _{\shortmid }\overline{\mathcal{\mathring{Z}}}%
^{\underline{P}\mu ^{\prime }\nu ^{\prime }}dV(u).  \label{act1}
\end{equation}%
In order to quantize the action following the path integral method, we have
to fix the ''gauge'' in order to suppress all equivalent field and
nonholonomic configurations related by nonholonomic/nonlinear gauge
transform, i.e. to introduce a gauge--fixing function $\mathcal{G}^{%
\underline{S}}(\mathbf{\mathring{Z}}_{\mu ^{\prime }}^{\underline{P}}),$ and
consider a generating source $J_{\mu ^{\prime }}^{\underline{P}}.$ The
partition function is taken%
\begin{eqnarray*}
Z[J] &=&\int [d\ _{\shortmid }\overline{\mathcal{\mathring{Z}}}]\det \left[
\frac{\delta \mathcal{G}^{\underline{S}}(\ ^{w}\mathbf{\mathring{Z}})}{%
\delta w^{\underline{P}}}\right] _{w=0}\times \\
&&\exp \left[ -\ _{\shortmid }\mathcal{\mathring{S}}(\mathbf{\mathring{Z}})-%
\frac{1}{2\alpha }\int \mathcal{G}^{\underline{S}}\mathcal{G}^{\underline{S}%
}dV(u)+J_{\mu ^{\prime }}^{\underline{P}}\mathbf{\mathring{Z}}_{\mu ^{\prime
}}^{\underline{P}}\right] ,
\end{eqnarray*}%
for $\alpha =const.$ Choosing the value $\mathcal{G}^{\underline{S}}=\
^{\circ }\widehat{\mathbf{D}}^{\nu ^{\prime }}\mathbf{\mathring{Z}}_{\nu
^{\prime }}^{\underline{S}},$ when the N--adapted infinitesimal nonlinear
gauge transform are approximated%
\begin{eqnarray*}
\mathbf{\mathring{Z}}_{\mu ^{\prime }}^{\underline{P}} &\rightarrow &\mathbf{%
\mathring{Z}}_{\mu ^{\prime }}^{\underline{P}}+\varkappa ^{-1}\ ^{\circ }%
\widehat{\mathbf{D}}_{\mu ^{\prime }}w^{\underline{P}}-f^{\underline{P}%
\underline{S}\underline{T}}w^{\underline{S}}\mathbf{\mathring{Z}}_{\mu
^{\prime }}^{\underline{T}}+\mathcal{O}(w^{2}) \\
&=&\mathbf{\mathring{Z}}_{\mu ^{\prime }}^{\underline{P}}+\varkappa
^{-1}\left( \ ^{\circ }\mathcal{D}_{\nu ^{\prime }}w\right) ^{\underline{P}}+%
\mathcal{O}(w^{2}),
\end{eqnarray*}%
and writing the determinant in terms of ghost (anticommutative variables) \
fields $\eta ^{\underline{P}},$ and their complex conjugated values $%
\overline{\eta }^{\underline{P}},$ we get the gravitational gauge Lagrangian%
\begin{equation}
\mathcal{L}=\frac{1}{4}\ \ _{\shortmid }\overline{\mathcal{\mathring{Z}}}_{\
\ \mu ^{\prime }\nu ^{\prime }}^{\underline{P}}\ \ _{\shortmid }\overline{%
\mathcal{\mathring{Z}}}^{\underline{P}\mu ^{\prime }\nu ^{\prime }}+\frac{1}{%
2\alpha }\left( \ ^{\circ }\widehat{\mathbf{D}}^{\nu ^{\prime }}\mathbf{%
\mathring{Z}}_{\nu ^{\prime }}^{\underline{S}}\right) \left( \ \ ^{\circ }%
\widehat{\mathbf{D}}^{\mu ^{\prime }}\mathbf{\mathring{Z}}_{\mu ^{\prime }}^{%
\underline{S}}\right) +\left( \ ^{\circ }\widehat{\mathbf{D}}^{\mu ^{\prime
}}\overline{\eta }^{\underline{P}}\right) \left( \ ^{\circ }\mathcal{D}_{\mu
^{\prime }}\eta \right) ^{\underline{P}}  \label{lagr2}
\end{equation}%
resulting respectively in propagators for the distortion and ghost fields,%
\begin{eqnarray*}
\left\langle \mathbf{\mathring{Z}}_{\mu ^{\prime }}^{\underline{P}}(\ ^{1}u)%
\mathbf{\mathring{Z}}_{\nu ^{\prime }}^{\underline{S}}(\ ^{2}u)\right\rangle
&=&\delta _{\mu ^{\prime }\nu ^{\prime }}\delta ^{\underline{P}\underline{S}%
}\bigtriangleup (\ ^{1}u-\ ^{2}u) \\
\mbox{\ and \ }\left\langle \eta ^{\underline{P}}(\ ^{1}u)\ \overline{\eta }%
^{\underline{P}}(\ ^{2}u)\right\rangle &=&\delta ^{\underline{P}\underline{S}%
}\bigtriangleup (\ ^{1}u-\ ^{2}u).
\end{eqnarray*}

We conclude that in our approach, the distortion gravitational field
(parametrized by de Sitter valued potentials) can be quantized similarly to
usual Yang--Mills fields but with nonholonomic/nonlinear gauge transforms,
all defined on a N--anholonomic base spacetime enabled with fundamental
geometric structures $\mathring{g}_{\alpha \beta },\ ^{\circ }\widehat{%
\mathbf{D}}_{\mu ^{\prime }}$ and $\mathbf{\mathring{e}}^{\nu ^{\prime }}.$

\subsection{N--adapted background field method}

The first examples of N--adapted background calculus were presented in Refs. %
\cite{vstr1,vstr2}\ when locally anisotropic (super) gravity configurations
were derived in low energy limits, with nonholonomic backgrounds, of (super)
string theory. In this section, we apply that formalism for quantization of
nonholonomic gauge gravity models.

\subsubsection{Splitting of nonholonomic gauge distortion fields}

We begin with a splitting if the nonholonomic gauge field into two parts $%
\mathbf{\mathring{Z}}_{\mu ^{\prime }}^{\underline{P}}\rightarrow \mathbf{%
\mathring{Z}}_{\mu ^{\prime }}^{\underline{P}}+\mathbf{B}_{\mu ^{\prime }}^{%
\underline{P}},$ where $\mathbf{\mathring{Z}}_{\mu ^{\prime }}^{\underline{P}%
}$ and $\mathbf{B}_{\mu ^{\prime }}^{\underline{P}}$ are called respectively
the quantum and background fields. The action $_{\shortmid }\mathcal{%
\mathring{S}}(\mathbf{\mathring{Z}}+\mathbf{B})$ (\ref{act1}) is invariant
under 1) quantum transforms%
\begin{eqnarray}
\delta \mathbf{\mathring{Z}}_{\mu ^{\prime }}^{\underline{P}} &=&\varkappa
^{-1}\left( \ ^{\circ }\widehat{\mathbf{D}}_{\mu ^{\prime }}w^{\underline{P}%
}+\varkappa f^{\underline{P}\underline{S}\underline{T}}w^{\underline{T}}%
\mathbf{B}_{\mu ^{\prime }}^{\underline{S}}\right) +f^{\underline{P}%
\underline{S}\underline{T}}w^{\underline{T}}\mathbf{\mathring{Z}}_{\mu
^{\prime }}^{\underline{S}}  \notag \\
&=&\varkappa ^{-1}\left( \ ^{B}\mathcal{D}_{\mu ^{\prime }}w\right) ^{%
\underline{P}}+f^{\underline{P}\underline{S}\underline{T}}w^{\underline{S}}%
\mathbf{\mathring{Z}}_{\mu ^{\prime }}^{\underline{T}}  \label{aux31} \\
\delta \mathbf{B}_{\mu ^{\prime }}^{\underline{P}} &=&0,  \notag
\end{eqnarray}%
and 2) background transforms%
\begin{eqnarray*}
\delta \mathbf{\mathring{Z}}_{\mu ^{\prime }}^{\underline{P}} &=&f^{%
\underline{P}\underline{S}\underline{T}}w^{\underline{T}}\mathbf{\mathring{Z}%
}_{\mu ^{\prime }}^{\underline{S}}, \\
\delta \mathbf{B}_{\mu ^{\prime }}^{\underline{P}} &=&\varkappa ^{-1}\
^{\circ }\widehat{\mathbf{D}}_{\mu ^{\prime }}w^{\underline{P}}+f^{%
\underline{P}\underline{S}\underline{T}}w^{\underline{T}}\mathbf{B}_{\mu
^{\prime }}^{\underline{S}}.
\end{eqnarray*}%
Following standard methods of quantization of gauge fields (see, for
instance, \cite{weinb2}), we derive from $Z[J]$ the partition function
\begin{eqnarray*}
&&Z[\mathbf{B}]=\int e^{-\mathcal{\mathring{S}}(\mathbf{\mathring{Z}},%
\mathbf{B})}[d\mathbf{\mathring{Z}}dcd\overline{c}]=\int \exp \{-\
_{\shortmid }\mathcal{\mathring{S}}(\mathbf{\mathring{Z}}+\mathbf{B})+ \\
&&tr\int \left( -\frac{1}{2\alpha }\left( \ ^{B}\mathcal{D}_{\mu ^{\prime }}%
\mathbf{\mathring{Z}}_{\mu ^{\prime }}\right) ^{2}+\overline{c}[\ ^{B}%
\mathcal{D}_{\mu ^{\prime }},\ \mathcal{D}_{\mu ^{\prime }}]c\right)
dV(u)\}[d\mathbf{\mathring{Z}}dcd\overline{c}],
\end{eqnarray*}%
with
\begin{equation}
\left( \mathcal{D}_{\mu ^{\prime }}w\right) ^{\underline{P}}=\ ^{\circ }%
\widehat{\mathbf{D}}_{\mu ^{\prime }}w^{\underline{P}}+\varkappa f^{%
\underline{P}\underline{S}\underline{T}}w^{\underline{T}}\left( \mathbf{%
\mathring{Z}}_{\mu ^{\prime }}^{\underline{S}}+\mathbf{B}_{\mu ^{\prime }}^{%
\underline{S}}\right)  \label{aux32}
\end{equation}%
and $\left( c,\overline{c}\right) $ being the Faddeev--Popov ghost fields.

One of the most important consequences of the background field method is
that the renormalization of the gauge constant, $\ ^{0}\varkappa =\
^{c}z\varkappa ,$ and the background field, $\ ^{0}\mathbf{B=\ }^{b}z^{1/2}%
\mathbf{B}$ are related,%
\begin{equation}
\ ^{c}z=\mathbf{\ }^{b}z^{-1/2},  \label{renc}
\end{equation}%
with renormalization of strength field,%
\begin{equation*}
\ ^{0}\mathbf{F}_{\mu \nu }^{\underline{S}}=\mathbf{\ }^{b}z^{1/2}\left[ \
^{\circ }\widehat{\mathbf{D}}_{\mu ^{\prime }}\mathbf{B}_{\nu ^{\prime }}^{%
\underline{S}}-\ ^{\circ }\widehat{\mathbf{D}}_{\nu ^{\prime }}\mathbf{B}%
_{\mu ^{\prime }}^{\underline{S}}+\varkappa \ ^{c}z\mathbf{\ }^{b}z^{1/2}f^{%
\underline{S}\underline{P}\underline{T}}\mathbf{B}_{\mu ^{\prime }}^{%
\underline{P}}\mathbf{B}_{\nu ^{\prime }}^{\underline{T}}\right] .
\end{equation*}%
This value defines the N--adapted background renormalization of $\
_{\shortmid }\overline{\mathcal{\mathring{Z}}}_{\ \ \mu ^{\prime }\nu
^{\prime }}^{\underline{P}}.$ One should be emphasized the substantial
physical difference between relations of type (\ref{renc}) in usual
Yang--Mills theory and nonholonomic gauge models of gravity. In the first
case, they relate renormalization of the fundamental constant (in
particular, electric charge) to renormalization of the background potential.
In the second case, we have certain constants fixing a d--connection, and a
lift in the total bundle, which must be renormalized in a compatible form
with renormalization of distortion gravitational fields. We can impose any
type of nonholonomic constraints on such nonlinear gauge fields but quantum
fluctuations re--define them and correlate to renormalization of
nonholonomic background configuration.

\subsubsection{Nonholonomic background effective action}

Using splitting $\mathbf{\mathring{Z}}_{\mu ^{\prime }}^{\underline{P}%
}\rightarrow \mathbf{\mathring{Z}}_{\mu ^{\prime }}^{\underline{P}}+\mathbf{B%
}_{\mu ^{\prime }}^{\underline{P}}$ and covariant derivatives $\ ^{B}%
\mathcal{D}_{\mu ^{\prime }}$ and $\mathcal{D}_{\mu ^{\prime }},$ see
respective formulas (\ref{aux31}) and (\ref{aux32}), we write the Lagrangian
(\ref{lagr2}) in the form%
\begin{equation}
\mathcal{L}=\frac{1}{4}\ _{\shortmid }\overline{\mathcal{\mathring{Z}}}_{\
\ \mu ^{\prime }\nu ^{\prime }}^{\underline{P}}\ _{\shortmid }\overline{%
\mathcal{\mathring{Z}}}^{\underline{P}\mu ^{\prime }\nu ^{\prime }}+\frac{1}{%
2\alpha }\left(\ ^{B}\mathcal{D}^{\nu ^{\prime }}\mathbf{\mathring{Z}}%
_{\nu ^{\prime }}^{\underline{S}}\right) \left( \ ^{B}\mathcal{D}^{\mu
^{\prime }}\mathbf{\mathring{Z}}_{\mu ^{\prime }}^{\underline{S}}\right)
+\left(\ ^{B}\mathcal{D}^{\mu ^{\prime }}\overline{\eta }^{\underline{P}%
}\right) \left(\mathcal{D}_{\mu ^{\prime }}\eta \right) ^{\underline{P}},
\label{lagr3}
\end{equation}%
where the total space curvature (field strength) depends both on quantum and
background fields, $\ _{\shortmid }\overline{\mathcal{\mathring{Z}}}_{\ \
\mu ^{\prime }\nu ^{\prime }}^{\underline{P}}=\ _{\shortmid }\overline{%
\mathcal{\mathring{Z}}}_{\ \ \mu ^{\prime }\nu ^{\prime }}^{\underline{P}%
}\left( \mathbf{\mathring{Z}},\mathbf{B}\right) .$ This Lagrangian is
similar to that used of standard Yang--Mills fields with chosen (for
convenience) Feynman gauge $\left( \alpha =1\right) .$ There are formal
geometric and substantial physical differences because we use the N--adapted
covariant operators $\ ^{\circ }\widehat{\mathbf{D}}_{\mu ^{\prime }}$
instead of partial derivatives $\partial _{\mu },$ our structure group is
the de Sitter group and constants in the theory do not characterize certain
fundamental gauge interactions but corresponding classes of nonholonomic
configurations. Nevertheless, a very similar Feynman diagram techniques can
be applied and a formal renormalization can be performed following standard
methods from quantum gauge fields theory (see, for instance, \cite%
{weinb2,freed,latorre,seijas,avd,smirnov,aguila,seij1}). For simplicity, we
shall omit details on explicit computations of diagrams but present the main
formulas and results and discuss the most important features of quantized
distortion fields in one-- and two--loop approximations.

The effective action for nonholonomic background derived for Lagrangian (\ref%
{lagr3}) can be written:
\begin{eqnarray}
^{eff}\Gamma \lbrack \mathbf{B}] &=&\frac{1}{2}\int \{\mathbf{B}_{\mu
^{\prime }}^{\underline{S}}(\ ^{1}u)\ ^{BB}\Gamma _{\mu ^{\prime }\nu
^{\prime }}^{\underline{S}\underline{T}}\left( \ ^{1}u-\ ^{2}u\right) \times
\notag \\
&&\mathbf{B}_{\nu ^{\prime }}^{\underline{T}}(\ ^{2}u)\ dV(\ ^{1}u)dV(\
^{2}u)\}+...  \label{efact3} \\
&=&\frac{1}{2}\int \mathbf{B}_{\mu ^{\prime }}^{\underline{S}}(\
^{1}u)\{\delta ^{\underline{S}\underline{T}}\ (\ ^{\circ }\widehat{\mathbf{D}%
}_{\mu ^{\prime }}\ ^{\circ }\widehat{\mathbf{D}}_{\nu ^{\prime }}-\delta
_{\mu ^{\prime }\nu ^{\prime }}\mathring{\square})\delta ^{(4)}(\ ^{1}u-\
^{2}u)  \notag \\
&&-\ _{\xi }^{BB}\Pi _{\mu ^{\prime }\nu ^{\prime }}^{\underline{S}%
\underline{T}}\left( \ ^{1}u-\ ^{2}u\right) \}\mathbf{B}_{\nu ^{\prime }}^{%
\underline{T}}(\ ^{2}u)\ dV(\ ^{1}u)dV(\ ^{2}u)\}+...  \notag \\
&=&\ ^{0}S[\mathbf{B}]+\ _{\xi }^{1}\Gamma -\frac{1}{2}\int \mathbf{B}_{\mu
^{\prime }}^{\underline{S}}(\ ^{1}u)\ _{\xi }^{BB}\Pi _{\mu ^{\prime }\nu
^{\prime }}^{\underline{S}\underline{T}}\left( \ ^{1}u-\ ^{2}u\right) \times
\notag \\
&&\mathbf{B}_{\nu ^{\prime }}^{\underline{T}}(\ ^{2}u)\ dV(\ ^{1}u)dV(\
^{2}u)+...,  \notag
\end{eqnarray}%
where $\ ^{0}S[\mathbf{B}]$ is called the three--level background two--point
function and $\xi =\alpha ^{-1}-1$ (in the Feynman gauge, we have $\xi =0).$
In the next sections, we will compute the one--loop contribution to the
background self--energy in this gauge. Using functional methods, we shall
also compute the term $\ _{\xi }^{1}\Gamma $ after expanding the complete
effective action at one loop at second order in the background fields and
collecting only components being linear in $\xi .$ Applying this procedure
to the remormalization group equation, we will take derivatives with respect
to parameter $\xi $ and impose, finally, the gauge $\xi =0.$

\subsubsection{One-- and two--loop computations}

To find the one--loop beta function we need to compute the background
self--energy. At the first step, we present the result for a renormalized
correction to the $\mathbf{B}_{\mu ^{\prime }}^{\underline{S}}$ propagator
(a similar calculus, for holonomic gauge fields, is provided in details in
section 3.1.2 of \cite{seij1}):%
\begin{eqnarray}
\ ^{\mathring{R}}\left\lfloor \ \ _{1-loop}^{BB}\Pi _{\mu ^{\prime }\nu
^{\prime }}^{\underline{S}\underline{T}}\left( u\right) \ \right\rfloor
&=&\varkappa ^{2}\ ^{1}C\delta ^{\underline{S}\underline{T}}\ (\ ^{\circ }%
\widehat{\mathbf{D}}_{\mu ^{\prime }}\ ^{\circ }\widehat{\mathbf{D}}_{\nu
^{\prime }}-\delta _{\mu ^{\prime }\nu ^{\prime }}\mathring{\square})\times
\label{aux41} \\
&&\left[ -\frac{11}{12(4\pi ^{4})}\mathring{\square}\frac{\mathring{l}n\left[
u^{2}\ \varpi ^{2}\right] }{u^{2}}-\frac{1}{72\pi ^{2}}\delta (u)\right] .
\notag
\end{eqnarray}%
This value is contained in effective action (\ref{efact3}).

In order to obtain an exact one--loop effective action we have to consider the part
of Lagrangian (\ref{lagr3}) which is quadratic on quantum distortion fields $%
\mathbf{\mathring{Z}}_{\mu ^{\prime }}^{\underline{P}}.$ Such an effective
action can be represented
\begin{eqnarray*}
\ ^{2}\mathcal{L} &=&\varkappa f^{\underline{P}\underline{S}\underline{T}}%
\mathcal{B}_{\mu ^{\prime }\nu ^{\prime }}^{\underline{P}}\mathbf{\mathring{Z%
}}_{\mu ^{\prime }}^{\underline{S}}\mathbf{\mathring{Z}}_{\nu ^{\prime }}^{%
\underline{T}}+\frac{1}{2}\left( \ \ ^{B}\mathcal{D}_{\mu ^{\prime }}\mathbf{%
\mathring{Z}}_{\nu ^{\prime }}\right) ^{\underline{S}}\left( \ ^{B}\mathcal{D%
}^{\mu ^{\prime }}\mathbf{\mathring{Z}}_{\mu ^{\prime }}\right) ^{\underline{%
S}} \\
&&+\frac{\xi }{2}\left( \ \ ^{B}\mathcal{D}^{\nu ^{\prime }}\mathbf{%
\mathring{Z}}_{\nu ^{\prime }}\right) ^{\underline{S}}\left( \ ^{B}\mathcal{D%
}^{\mu ^{\prime }}\mathbf{\mathring{Z}}_{\mu ^{\prime }}\right) ^{\underline{%
S}} \\
&=&-\frac{1}{2}\mathbf{\mathring{Z}}_{\mu ^{\prime }}^{\underline{S}}\left[
\delta _{\mu ^{\prime }\nu ^{\prime }}\breve{\square}^{\underline{S}%
\underline{T}}-2\varkappa f^{\underline{P}\underline{S}\underline{T}}%
\mathcal{B}_{\mu ^{\prime }\nu ^{\prime }}^{\underline{P}}+\xi (\ \ ^{B}%
\mathcal{D}_{\mu ^{\prime }}\ ^{B}\mathcal{D}_{\nu ^{\prime }})^{\underline{S%
}\underline{T}}\right] \mathbf{\mathring{Z}}_{\nu ^{\prime }}^{\underline{T}%
},
\end{eqnarray*}%
where $\breve{\square}^{\underline{S}\underline{T}}=(\ \ ^{B}\mathcal{D}%
_{\mu ^{\prime }}\ ^{B}\mathcal{D}^{\mu ^{\prime }})^{\underline{S}%
\underline{T}}$ and
\begin{equation*}
\mathcal{B}_{\mu ^{\prime }\nu ^{\prime }}^{\underline{P}}=\ ^{\circ }%
\widehat{\mathbf{D}}_{\mu ^{\prime }}\mathbf{B}_{\nu ^{\prime }}^{\underline{%
P}}-\ ^{\circ }\widehat{\mathbf{D}}_{\nu ^{\prime }}\mathbf{B}_{\mu ^{\prime
}}^{\underline{P}}+\varkappa f^{\underline{P}\underline{S}\underline{T}}%
\mathbf{B}_{\mu ^{\prime }}^{\underline{S}}\mathbf{B}_{\nu ^{\prime }}^{%
\underline{T}}.
\end{equation*}%
The terms from $[...]$ define the generated functional $\mathcal{G}$ for
connected Green functions,
\begin{eqnarray}
\mathcal{G} &=&-tr\ln \left[ \delta _{\mu ^{\prime }\nu ^{\prime }}\breve{%
\square}^{\underline{S}\underline{T}}-2\varkappa f^{\underline{P}\underline{S%
}\underline{T}}\mathcal{B}_{\mu ^{\prime }\nu ^{\prime }}^{\underline{P}%
}+\xi (\ \ ^{B}\mathcal{D}_{\mu ^{\prime }}\ ^{B}\mathcal{D}_{\nu ^{\prime
}})^{\underline{S}\underline{T}}\right]  \label{green1} \\
&\simeq &\mathcal{G}_{0}+\xi \ ^{1}C\varkappa ^{2}tr[\frac{1}{2}\mathring{%
\bigtriangleup}\mathcal{B}_{\mu ^{\prime }\nu ^{\prime }}^{\underline{P}}%
\mathring{\bigtriangleup}\mathcal{B}_{\mu ^{\prime }\nu ^{\prime }}^{%
\underline{P}}  \notag \\
&&-2\mathring{\bigtriangleup}\mathcal{B}_{\mu ^{\prime }\nu ^{\prime }}^{%
\underline{P}}\mathring{\bigtriangleup}\mathcal{B}_{\nu ^{\prime }\lambda
^{\prime }}^{\underline{P}}\mathring{\bigtriangleup}\left( \ ^{\circ }%
\widehat{\mathbf{D}}_{\lambda ^{\prime }}\ ^{\circ }\widehat{\mathbf{D}}%
_{\nu ^{\prime }}\right) ]+O(\xi ^{2},\mathbf{B}^{3}),  \notag
\end{eqnarray}%
where $\mathring{\bigtriangleup}=\mathring{\square}^{-1},$ for $\mathring{%
\square}=\ ^{\circ }\widehat{\mathbf{D}}^{\nu ^{\prime }}\ ^{\circ }\widehat{%
\mathbf{D}}_{\nu ^{\prime }}.$

The renormalized version of $\mathcal{G}$ (\ref{green1}) gives the term used
in effective action (\ref{efact3}),%
\begin{eqnarray}
\ _{\xi }^{1}\Gamma &=&-\frac{\xi \ ^{1}C\varkappa ^{2}}{16\pi ^{2}}\int
\mathbf{B}_{\mu ^{\prime }}^{\underline{S}}(\ ^{1}u)\mathbf{B}_{\nu ^{\prime
}}^{\underline{T}}(\ ^{2}u)\left( \ _{\ ^{1}u}^{\circ }\widehat{\mathbf{D}}%
_{\mu ^{\prime }}\ _{\ ^{1}u}^{\circ }\widehat{\mathbf{D}}_{\nu ^{\prime
}}-\delta _{\mu ^{\prime }\nu ^{\prime }}\mathring{\square}\right)  \notag \\
&&\left( \mathring{\square}\bigtriangleup (\ ^{1}u-\ ^{2}u)\right) dV(\
^{1}u)dV(\ ^{2}u),  \label{aux42}
\end{eqnarray}%
were, for instance, $\ _{\ ^{1}u}^{\circ }\widehat{\mathbf{D}}_{\mu ^{\prime
}}$ denotes that operator $\ _{\ }^{\circ }\widehat{\mathbf{D}}_{\mu
^{\prime }}$ is computed in point $^{1}u.$

Now we present the formula for the two--loop renormalized contribution to
the background field self--energy (it is a result of cumbersome nonholonomic
background computations using $H$--components from Appendix \ref{asovd}, see
similar computations with Feynman diagrams in section 3.1.3 of \cite{seij1}).%
\begin{equation}
\ ^{BB}_{(2-loop)}\Pi _{\mu ^{\prime }\nu ^{\prime }}^{\underline{S}%
\underline{T}}\left( u\right) =-\frac{\ ^{1}C\varkappa ^{4}}{128\pi ^{2}}%
\delta ^{\underline{S}\underline{T}}\left( \ ^{\circ }\widehat{\mathbf{D}}%
_{\mu ^{\prime }}\ ^{\circ }\widehat{\mathbf{D}}_{\nu ^{\prime }}-\delta
_{\mu ^{\prime }\nu ^{\prime }}\mathring{\square}\right) \mathring{\square}%
\frac{\mathring{l}n\left[ u^{2}\ \varpi ^{2}\right] }{u^{2}}+...
\label{aux43}
\end{equation}%
We emphasize that applying the CDR$_{N}$ method for one--loop formulas we
fix a priori a renormalization scheme resulting in total two--loop
renormalized contributions to the background self--energy for distortion
gravitational field.

\subsubsection{Renormalization group equations for nonholonomic
configurations}

Let us parametrize
\begin{equation*}
^{BB}\Gamma _{\mu ^{\prime }\nu ^{\prime }}^{\underline{S}\underline{T}%
}\left( u\right) =\delta ^{\underline{S}\underline{T}}\left( \ ^{\circ }%
\widehat{\mathbf{D}}_{\mu ^{\prime }}\ ^{\circ }\widehat{\mathbf{D}}_{\nu
^{\prime }}-\delta _{\mu ^{\prime }\nu ^{\prime }}\mathring{\square}\right)
\Gamma ^{(2)}\left( u\right) ,
\end{equation*}%
where $\Gamma ^{(2)}(u)$ is computed as the sum of terms (\ref{aux41}), (\ref%
{aux42}) and (\ref{aux43}), i.e.%
\begin{equation}
\Gamma ^{(2)}=\left[\frac{1}{\varkappa ^{2}}+\left(\frac{1}{9}+\xi
\right) \frac{\ ^{1}C}{8\pi ^{2}}\right] \delta (u)+\frac{\ ^{1}C}{2(2\pi
)^{4}}\left( \frac{11}{6}+\frac{\ ^{1}C\varkappa ^{2}}{(2\pi )^{2}}\right)
\mathring{\square}\frac{\mathring{l}n\left[u^{2}\ \varpi ^{2}\right] }{u^{2}%
}...  \label{aux44}
\end{equation}%
For this function, we consider the renormalization group (RG) equation%
\begin{equation*}
\left[ \varpi \frac{\partial }{\partial \varpi }+\beta (\varkappa )\frac{%
\partial }{\partial \varkappa }+\ _{\xi }\gamma \frac{\partial }{\partial
\xi }-2\ _{B}\gamma \right] \Gamma ^{(2)}\left( u\right) =0
\end{equation*}%
when $_{\xi }\gamma =\ -5\ ^{1}C\varkappa ^{2}/24\pi ^{2}.$ We can transform
$\ _{B}\gamma =0$ if the background gauge gravity field is redefined $%
B^{\prime }=\varkappa B$ for the charge and background field
renormalizations being related by formula (\ref{renc}).

We can evaluate the first two coefficients, $\ _{1}\beta $ and $\ _{2}\beta
, $ of the expansion of the beta function%
\begin{equation*}
\beta (\varkappa )=\ _{1}\beta \varkappa ^{3}+\ _{2}\beta \varkappa ^{5}+%
\mathcal{O}(\varkappa ^{7}),
\end{equation*}%
with
\begin{equation*}
\ _{1}\beta =-\frac{11\ ^{1}C}{48\pi ^{2}}\mbox{ \ and \ }\ _{2}\beta =-%
\frac{17\left( \ ^{1}C\right) ^{2}}{24(2\pi )^{4}}.
\end{equation*}%
These formulas are similar to those in (super) Yang--Mills theory (see for
instance \cite{jones,ferrara}). For the nonholonomic gauge model of the
Einstein gravity, such coefficients are not related to renormalization of a
fundamental gauge constant but to quantum redefinition of certain constants
stating a nonholonomic configuration for gravitational distortion fields.

\section{Summary and Conclusions}

In the present paper, we have elaborated a (one-- and two--loop) perturbative
quantization approach to Einstein gravity using a two--connection
formalism and nonholonomic gauge gravity models. The essential technique is
the use of the geometry of nonholonomic distributions and adapted frames and
(non) linear connections which are completely defined by a given (pseudo) Riemannian metric tensor.
\vskip5pt

Let us outline the key steps of the quantization algorithm we developed:

\begin{enumerate}
\item On a four dimensional pseudo--Riemannian spacetime manifold $\mathbf{V,%
}$ we can consider any distribution of geometric objects, frames and local
coordinates. For our purposes, there were involved nonholonomic
distributions inducing $(2+2)$--dimensional spacetime splitting
characterized by corresponding classes of nonholonomic frames and associated
nonlinear connection (N--connection) structures.

\item We applied the formalism of N--connections and distinguished connections (d--connections) completely
determined by metric structure. This allows us to rewrite equivalently the
Einstein equations in terms of nonholonomic variables (vierbein fields and
generalized connections). Such  geometric formulations of Einstein gravity
  are more suitable for quantization following the background field
method and techniques elaborated in Yang--Mills theory.

\item The background field method can be modified for d--connections. It is
convenient to construct such a canonical d--connection which is characterized,
for instance, by certain constant matrix coefficients of the Riemannian and
Ricci tensors. For an auxiliary model of gravity with d--connections and
infinitely many couplings, we impose such nonholonomic constraints when the
Ricci and Riemannian/Weyl tensors vanish (nevertheless, similar tensors
corresponding to the Levi--Civita connection are not trivial). Such a model
can be quantized and renormalized as a gravity theory with two constants,
for instance, with the same gravitational and cosmological constants as in
Einstein gravity.

\item Any background d--connection completely defined by a metric structure
can be distorted in a unique form to the corresponding Levi--Civita
connection. The distortion field can be encoded into a class of Yang--Mills like
equations for nonholonomic gauge gravity models. The constants in
such theories do not characterize any additional gauge like gravitational
interactions but certain constraints imposed on the nonholonomic structure
which give us the possibility to establish equivalence with the Einstein
equations on the base spaceime.

\item The nonholonomic gauge gravity model can be quantized following
methods elaborated for Yang--Mills  fields. There are also some
important differences: We work on background spaces enabled with  d--connections with constant curvature
coefficients (and quantized as in point 3); gravitational gauge group transforms are generic nonlinear and
nonholonomically deformed, the constants in the theory, and their renorm
group flows, are not related to certain additional gravitational--gauge
interactions but determined by self--consistent quantum flows of nonholonomic
configurations.

\item Above outlined  geometric and quantum constructions
can be redefined in terms of the Levi--Civita connection and corresponding
Einstein equations. Renormalized d--connection and distortion fields and
redefined gravitational, cosmological and nonholonomic configuration
constants will be regrouped correspondingly for the metric/ tetradic
components.
\end{enumerate}

The proposed quantization algorithm is based on three important geometric
ideas:

The first one is that for a metric tensor we can construct an infinite number of metric compatible linear connections (in N--connection adapted form, d--connections). Even the torsion of such a d--connection is not zero, it can be considered as a nonholonomic frame effect with coefficients induced
completely by certain off--diagonal coefficients of the metric tensor. Using two linear connections completely defined by the same metric structure, we get more possibilities in approaching  the problem of renormalizability
of gravity. We can invert equivalently all construction in terms of the Levi--Civita connection and work in certain "standard" variables of general relativity theory. It is not obligatory to generalize the Einstein gravity theory to some  models of
Einstein--Cartan/ string/ gauge gravity, were the torsion fields are subjected to satisfy certain additional field (dynamical, or algebraic) equations.

The second geometric idea is to consider such nonholonomic distributions
when the so--called canonical d--connection is characterized by constant
matrix coefficients of Riemannian and Ricci tensors. We can prescribe such
sets of nonholonomic constants when the one-- and two--loop quantum
divergent terms vanish. The nonrenormalizability of Einstein's theory\ (in a
standard meaning of gauge theories with certain gauge symmetry and mass
dimensionality of couplings) is related to the fact that the gravitational
coupling is characterized by Newton's constant, which for a four dimensional
spacetime has the dimension of a negative power of mass.\footnote{ This resulted in
conclusion that the removal of divergences of quantum gravity is possible
only in the presence of infinitely many independent coupling constants \cite%
{anselmi}.}  The background field method can be redefined for d--connections
and nonholonomic configurations when certain models of quantum nonholonomic
gravity with infinite coupling constants transform into a theory with usual
gravitational and cosmological constants.

The third geometric idea is to construct models of gauge theories generated
from the Einstein gravity theory imposing a corresponding class of nonholonomic constraints. Certain
additional constants (prescribing a nonholonomic distribution)
characterize a nonholonomic field/ geometric configuration and do not involve any additional
(to gravity) interactions. This type of nonholonomically constrained gauge gravitational interactions are
completely defined by the components of a distortion tensor (also uniquely
defined by a metric tensor) from a chosen d--connection to the Levi--Civita
connection. For such models, we can perform quantization and apply formal
renormalization schemes following standard methods of gauge theories. In
this work we developed the method of differential renormalization for
nonholonomic backgrounds and gauge like distortion (gravitational) fields.

The above presented geometric ideas and quantization procedure for the Einstein gravity theory in nonholonomic variables could be viewed as starting point for a perturbative approach  relating our former results on nonholonomic (and nonperturbative) loop constructions, Fedosov--Lagrange--Hamilton quantization, nonholonomic string--brane quantum models of gravity and (non) commutative models of gauge gravity, see  \cite{vpla,vfqlf,vqg4,anav,vgwgr,vloopdq,vggr,vncg,vgon,vncg1,vd} and references therein.

\vskip5pt

\textbf{Acknowledgement: }A part of this work was partially performed during
a visit at the Fields Institute.

\setcounter{equation}{0} \renewcommand{\theequation}
{A.\arabic{equation}} \setcounter{subsection}{0}
\renewcommand{\thesubsection}
{A.\arabic{subsection}}

\appendix

\section{Overlapping Divergences on Nonholonomic \newline  Spa\-ces}

\label{asovd}In this appendix, present some explicit formulas for
overlapping divergences.%
\begin{eqnarray*}
\ \ \ ^{\mathring{R}}H[1,1;1,1] &=&a\bigtriangleup , \\
\ \ \ ^{\mathring{R}}H[\ ^{\circ }\widehat{\mathbf{D}}_{\mu ^{\prime
}},1;1,1] &=&\frac{a}{2}\ ^{\circ }\widehat{\mathbf{D}}_{\mu ^{\prime
}}\bigtriangleup ,\mbox{\ for \ }a=const, \\
\ \ ^{\mathring{R}}H[1,\ ^{\circ }\widehat{\mathbf{D}}_{\mu ^{\prime }},;1,\
^{\circ }\widehat{\mathbf{D}}^{\mu ^{\prime }}] &=&-\frac{1}{16(2\pi )^{6}}%
\mathring{\square}\frac{\mathring{l}n\left[ u^{2}\ \varpi ^{2}\right] }{u^{2}%
}+..., \\
\ ^{\circ }\widehat{\mathbf{D}}_{\lambda ^{\prime }}\ ^{\mathring{R}}H[1,\
^{\circ }\widehat{\mathbf{D}}_{\mu ^{\prime }};1,\ ^{\circ }\widehat{\mathbf{%
D}}^{\lambda ^{\prime }}] &=&-\frac{1}{64(2\pi )^{6}}\ ^{\circ }\widehat{%
\mathbf{D}}_{\mu ^{\prime }} \mathring{\square}\frac{\mathring{l}n\left[
u^{2}\ \varpi ^{2}\right] }{u^{2}}+...,
\end{eqnarray*}

\begin{eqnarray*}
\ ^{\circ }\widehat{\mathbf{D}}_{\lambda ^{\prime }}\ ^{\mathring{R}}H[1,1;\
^{\circ }\widehat{\mathbf{D}}^{\lambda ^{\prime }}\ ^{\circ }\widehat{%
\mathbf{D}}_{\mu ^{\prime }},1] &=&\frac{1}{128(2\pi )^{6}}\ ^{\circ }%
\widehat{\mathbf{D}}_{\mu ^{\prime }} \\
&&\mathring{\square}\frac{\left( \mathring{l}n\left[ u^{2}\ \varpi ^{2}%
\right] \right) ^{2}+3\mathring{l}n\left[ u^{2}\ \varpi ^{2}\right] }{u^{2}}%
+..., \\
\ ^{\mathring{R}}H[1,\ ^{\circ }\widehat{\mathbf{D}}_{\lambda ^{\prime }};\
^{\circ }\widehat{\mathbf{D}}^{\lambda ^{\prime }}\ ^{\circ }\widehat{%
\mathbf{D}}_{\mu ^{\prime }},1] &=&\frac{1}{256(2\pi )^{6}}\ ^{\circ }%
\widehat{\mathbf{D}}_{\mu ^{\prime }} \\
&&\mathring{\square}\frac{\left( \mathring{l}n\left[ u^{2}\ \varpi ^{2}%
\right] \right) ^{2}-7\mathring{l}n\left[ u^{2}\ \varpi ^{2}\right] }{u^{2}}%
+..., \\
\ ^{\mathring{R}}H[\ ^{\circ }\widehat{\mathbf{D}}_{\mu ^{\prime }}\ ^{\circ
}\widehat{\mathbf{D}}^{\lambda ^{\prime }},\ ^{\circ }\widehat{\mathbf{D}}%
_{\lambda ^{\prime }};1,1] &=&-\frac{1}{64(2\pi )^{6}}\ ^{\circ }\widehat{%
\mathbf{D}}_{\mu ^{\prime }} \\
&&\mathring{\square}\frac{\left( \mathring{l}n\left[ u^{2}\ \varpi ^{2}%
\right] \right) ^{2}+2\mathring{l}n\left[ u^{2}\ \varpi ^{2}\right] }{u^{2}}%
+...,
\end{eqnarray*}

\begin{eqnarray*}
\ ^{\circ }\widehat{\mathbf{D}}_{\lambda ^{\prime }}\ ^{\mathring{R}}H[1,\
^{\circ }\widehat{\mathbf{D}}_{\mu ^{\prime }};\ ^{\circ }\widehat{\mathbf{D}%
}_{\nu ^{\prime }}\ ^{\circ }\widehat{\mathbf{D}}^{\lambda ^{\prime }},1] &=&%
\frac{1}{256(2\pi )^{6}} [- 2\delta _{\mu ^{\prime }\nu ^{\prime }}\mathring{%
\square}^{2}\frac{\mathring{l}n\left[ u^{2}\ \varpi ^{2}\right] }{u^{2}} + \\
\ ^{\circ }\widehat{\mathbf{D}}_{\mu ^{\prime }}\ ^{\circ }\widehat{\mathbf{D%
}}_{\nu ^{\prime }}&&\frac{\left( \mathring{l}n\left[ u^{2}\ \varpi ^{2}%
\right] \right) ^{2}+\mathring{l}n\left[ u^{2}\ \varpi ^{2}\right] }{u^{2}}]
+ ..., \\
\ \ ^{\mathring{R}}H[1,\ ^{\circ }\widehat{\mathbf{D}}_{\mu ^{\prime }},;1,\
^{\circ }\widehat{\mathbf{D}}_{\nu ^{\prime }}] &=&-\frac{1}{64(2\pi )^{6}}%
\delta _{\mu ^{\prime }\nu ^{\prime }}\mathring{\square}\frac{\mathring{l}n%
\left[ u^{2}\ \varpi ^{2}\right] }{u^{2}}]+..., \\
&&...
\end{eqnarray*}%
For simplicity, we omit the rest of formulas which are similar to those
presented in Ref. \cite{seij1} (we have to substitute formally those
holonomic operators into nonholonomic covariant ones, for a corresponding
d--connection with constant coefficient curvature).

\end{document}